\documentclass[preprint,12pt]{elsarticle}

\usepackage{amssymb}
\usepackage{amsthm}

\usepackage{amsmath}
\usepackage{enumitem}
\usepackage{bm}
\usepackage{graphicx}
\usepackage{graphics}
\usepackage{subfigure}
\usepackage{subcaption}
\usepackage{tabularx,caption}
\usepackage{float}
\usepackage{color}
\usepackage{multirow}
\usepackage{booktabs}
\usepackage{diagbox}
\usepackage{makecell}
\usepackage{lineno}
\usepackage{tikz}
\usepackage{xcolor}
\usepackage[percent]{overpic}
\usepackage{pdfpages}
\usepackage{algorithm}
\usepackage{algpseudocode}
\usepackage{multirow}
\usepackage{graphicx}

\journal{XXX}

\begin{document}

\begin{frontmatter}

\title{Physics-informed sampler for solver-free data generation in operator learning}

\author[hust]{Wen You \fnref{1}}
\author[hust]{Shaoqian Zhou \fnref{1}}
\author[westlake]{Dixia Fan}
\author[hust]{Xuhui Meng \fnref{2}}

\address[hust]{Institute of Interdisciplinary Research for Mathematics and Applied Science,
  School of Mathematics and Statistics, Huazhong University of Science and Technology,
  Wuhan, China}
\address[westlake]{School of Engineering, Westlake University, Hangzhou, China}

\fntext[1]{The first two authors contributed equally to this work.}
\fntext[2]{Corresponding author: xuhui\_meng@hust.edu.cn (Xuhui Meng).}

\begin{abstract}
Neural operators provide fast surrogates for parametric partial differential equations (PDEs), but their training commonly requires large datasets generated by repeated numerical PDE solves. Efficient and scalable  methods for training data generation are therefore desirable.  In this work, we present a physics-informed sampler (PI-sampler) for generating PDE-consistent training data without repeated numerical PDE solves. Inspired by the stochastic solution priors used in Bayesian physics-informed neural networks, the PI-sampler draws candidate solution fields from prescribed neural-network priors and evaluates the governing differential, boundary, and initial-condition operators through automatic differentiation. Each draw thereby produces a consistent PDE-input--solution tuple by construction. In contrast to solution-first approaches based on analytical solution bases or precomputed seed solutions, the PI-sampler requires neither problem-specific basis functions nor prior numerical solutions. The construction in the PI-sampler defines an implicit pushforward distribution over PDE inputs. For linear differential operators, this distribution can be characterized under the Gaussian-process limit of the BNN; for nonlinear operators, it is generally non-Gaussian and analytically intractable but remains directly sampleable. We combine the PI-sampler with a function encoder and a Transformer-based operator learning architecture and evaluate the resulting framework on nonlinear reaction--diffusion, Poisson equations on parameterized geometries, time-dependent Sine--Gordon dynamics, a five-dimensional PDE, and vortex-induced vibration of a flexible cylinder. The resulting models accurately predict cases involving independently prescribed forcing distributions, homogeneous forcing, mixed boundary conditions, and previously unseen geometries. A DeepONet trained using the same reaction-diffusion samples further demonstrates that the generated data are not specific to the latent Transformer operator architecture. In the reported problem-specific comparisons, the PI-sampler generates data about 51 and 850 times faster than the corresponding numerical solvers when run on the CPU and GPU, respectively. Because the induced input distribution is controlled only indirectly through the solution prior for nonlinear PDEs, we envision the PI-sampler as an inexpensive mechanism for data generation and augmentation that complements high-fidelity simulations for operator learning.
\end{abstract}

\begin{keyword}
physics-informed sampler \sep operator learning \sep Bayesian neural networks
\sep pushforward measure \sep nonlinear PDEs
\end{keyword}

\end{frontmatter}


\section{Introduction}
Partial differential equations (PDEs) are fundamental models of physical
processes in science and engineering, including fluid flow, wave
propagation, heat transfer, and transport phenomena
\cite{schunk1977mathematical,zhang2019quantifying,mazumder2021boltzmann,lin2024monte}.
Finite-difference, finite-volume, and finite-element methods provide reliable
approximations to their solutions. Their cost, however, can become
prohibitive when the same PDE must be solved repeatedly under varying source
terms, boundary or initial conditions, physical parameters, or geometries,
as occurs in uncertainty quantification, inverse problems, optimization,
control, and real-time prediction \cite{li2020fourier,azizzadenesheli2024neural}.

Neural operators address this repeated-query setting by learning mappings
between function spaces \cite{lu2021learning,li2020fourier,kovachki2023neural}.
Once trained on paired input and solution fields, they can evaluate an
approximation to a PDE solution operator much faster than a conventional
solver. Representative architectures include deep operator networks (DeepONet)
\cite{lu2021learning,lu2022comprehensive,jin2022mionet}, Fourier neural operator (FNO)
\cite{li2020fourier,kovachki2023neural,li2023geometry}, Transformer-based
neural operators \cite{cao2021choose,li2023oformer,hao2023gnot,
wu2024transolver}, and kernel-based operator-learning methods
\cite{nelsen2021random,batlle2024kernel}. A conventional neural-operator
training pipeline first samples PDE inputs, such as forcing terms and
boundary or initial conditions, from prescribed distributions and then
invokes a numerical solver for every realization. Its cost therefore grows
with both the number of training
samples and the difficulty of each solve, and is especially expensive for
nonlinear, high-dimensional, time-dependent, or geometrically varying
problems. Physics-informed neural operators reduce dependence on labeled
solutions by incorporating PDE residuals into training
\cite{li2021physicsinformed}, but replace part of the data-generation burden
with a potentially challenging physics-constrained optimization problem.

One way to bypass repeated solves is to reverse the usual direction of data generation. The method of manufactured solutions provides a classical solution-first procedure
for numerical-code verification: an analytical field is chosen and
substituted into the governing equation to derive the corresponding source
term and boundary or initial conditions \cite{salari2000code}. Recent studies
have adapted this principle to neural-operator data generation. Hasani and Ward
\cite{hasani2025generating} sample candidate solutions as random linear
combinations of Laplace eigenfunctions and obtain the corresponding
right-hand sides by differentiation. Their construction provides a
principled basis for steady elliptic problems when the required
problem- and geometry-specific eigenfunctions are available. DiffOAS
\cite{dong2024accelerating} also applies differential operators in solution
space, using combinations of a small set of precomputed PDE solutions to
accelerate dataset construction. These works demonstrate that solution-space
generation can substantially reduce the cost of constructing training data.
They nevertheless retain problem-specific dependencies: the former requires
known eigenfunction bases, whereas the latter starts from seed solutions
computed by a PDE solver. These requirements can complicate their extension
across equations and geometries.


The same solution-first perspective also appears in Bayesian
physics-informed learning. Bayesian physics-informed neural networks place
prior distributions on neural-network parameters, thereby inducing a prior
over solution fields, and propagate sampled fields through the governing
operators \cite{yang2021b,meng2022learning}. The present work is motivated
directly by this construction but uses it for a different purpose. Rather
than conditioning the prior on observations for Bayesian inference, we use
unconditioned prior samples to construct paired data for operator learning.
Specifically, we develop a physics-informed sampler (PI-sampler) in which a
random neural network with prescribed parameter priors generates candidate
solution fields. Automatic differentiation applies the differential,
boundary, and initial-condition operators to each field, producing a
PDE-consistent input--solution tuple. Thus, repeated PDE solves are replaced
by sampling and differentiation. The PI-sampler requires neither an
analytical solution basis as in \cite{hasani2025generating} nor precomputed
seed solutions as in \cite{dong2024accelerating}. Its random neural-field
representation applies to steady and space-time domains, accommodates
isotropic or anisotropic prior scales, and extends naturally to
high-dimensional problems, due to the flexibility of neural networks. As we will discuss in the 
following section, the induced PDE-input distribution is controlled indirectly
through the solution prior for nonlinear PDEs.  The PI-sampler is hence intended to complement rather
than universally replace targeted solver-generated data.

Note that the PI-sampler outputs paired functional data, it is independent of the downstream operator architecture. For
the heterogeneous and high-dimensional examples considered here, we pair it
with a latent Transformer operator (LTO) comprising a function encoder and a
Transformer. The LTO is a supporting implementation rather than
a required component of the sampler; a multi-input DeepONet trained on the
same PI-sampler data in one benchmark provides an additional architecture
check.

{
The main contributions of this work are summarized as follows:

\begin{enumerate}

    \item We develop a physics-informed sampler based on random neural fields for generating PDE-consistent input--solution pairs without repeated numerical PDE solves. Unlike existing solution-first constructions based on analytical eigenfunction expansions \cite{hasani2025generating} or combinations of precomputed solutions \cite{dong2024accelerating}, the proposed sampler requires neither problem-specific solution bases nor solver-generated seed solutions. The same construction can therefore be instantiated for steady, time-dependent, nonlinear, high-dimensional, and geometrically varying problems.

 \item We interpret the sampler as an implicit pushforward model. The induced forcing process is analytically characterizable for linear operators under the GP limit and directly sampleable for nonlinear operators. Mixtures of BNN priors are used to broaden the represented functional scales. 
 
    \item We evaluate the method on nonlinear, geometrically varying, time-dependent, high-dimensional, and engineering problems. The experiments examine independent forcing distributions, homogeneous forcing, mixed boundary conditions, unseen geometries, and compatibility with different neural-operator architectures.

       \item A controlled Poisson comparison shows that target-aware FEM data are more sample-efficient, whereas the PI-sampler reduces data-generation time by approximately $51.2\times$ on the same CPU system and $852\times$ with GPU batching.


\end{enumerate}
}




The remainder of the paper is organized as follows. Section~2 introduces
the PI-sampler and its probabilistic interpretation. Section~3 presents the
LTO used in the numerical studies. Section~4 reports the results and
discusses the scope and limitations of the approach. Section~5 summarizes the
paper.

\section{Methodology}
\label{sec:pi_sampler}

{This section first formulates the operator-learning problem and the conventional solver-based data pipeline. It then introduces the PI-sampler and interprets its induced input distribution as a pushforward measure.}

\subsection{Problem formulation}
\label{sec:problem_formulation}

We consider a general parameterized partial differential equation of the
form
\begin{equation}
\mathcal{N}_{\bm{\beta}}[u](t,\boldsymbol{x})
=
f(t,\boldsymbol{x}),
\qquad
\boldsymbol{x}\in\Omega_{\boldsymbol{x}},
\quad
t\in\Omega_t,
\quad
\bm{\beta}\in\mathcal{P},
\label{eq:governing_equation}
\end{equation}
subject to the boundary condition
\begin{equation}
\mathcal{B}_{\bm{\beta}}[u]
(t,\boldsymbol{x}_{\mathrm{bc}})
=
b(t,\boldsymbol{x}_{\mathrm{bc}}),
\qquad
\boldsymbol{x}_{\mathrm{bc}}
\in
\partial\Omega_{\boldsymbol{x}},
\label{eq:boundary_condition}
\end{equation}
and, for time-dependent problems, the initial condition
\begin{equation}
u(0,\boldsymbol{x})
=
u_0(\boldsymbol{x}).
\label{eq:initial_condition}
\end{equation}
Here, $u$ denotes the solution field, $f$ is the source or forcing term,
$\mathcal{N}_{\bm{\beta}}$ is a possibly nonlinear differential operator
parameterized by $\bm{\beta}$, and $\mathcal{B}_{\bm{\beta}}$ denotes the boundary
operator. The spatial domain is denoted by
$\Omega_{\boldsymbol{x}}$, its boundary by
$\partial\Omega_{\boldsymbol{x}}$, and the temporal domain by $\Omega_t$.
Depending on the problem, the equation parameters, boundary and initial
conditions, forcing term, and computational geometry may vary among
different problem instances.

The corresponding solution operator can be written abstractly as
\[
\mathcal{G}:
\left(
f,
b,
u_0,
\bm{\beta},
\Omega_{\boldsymbol{x}}
\right)
\longmapsto
u.
\label{eq:solution_operator}
\]
Only a subset of the quantities on the left-hand side may vary in a
specific application. For example, a steady equation on a fixed domain may
involve only the mapping $(f, b) \mapsto u$, whereas a more general problem may
involve source terms, boundary and initial conditions, equation parameters,
and geometries simultaneously.

In a conventional data-generation procedure, probability distributions are
first assigned to the PDE inputs:
\[
f^{(i)}
\sim
\mu_f,
\qquad
b^{(i)}
\sim
\mu_b,
\qquad
u_0^{(i)}
\sim
\mu_0,
\qquad
\beta^{(i)}
\sim
\mu_{\beta}.
\label{eq:conventional_input_sampling}
\]
{For each sampled input tuple, a numerical solver is then used to obtain the corresponding solution,}
\[
u^{(i)}
=
\mathcal{G}
\left(
f^{(i)},
b^{(i)},
u_0^{(i)},
\beta^{(i)},
\Omega_{\boldsymbol{x}}^{(i)}
\right).
\label{eq:conventional_solution_generation}
\]
Repeating this process produces a training dataset
\[
\mathcal{D}_{\mathrm{solver}}
=
\left\{
\left(
f^{(i)},
b^{(i)},
u_0^{(i)},
\beta^{(i)},
\Omega_{\boldsymbol{x}}^{(i)};
u^{(i)}
\right)
\right\}_{i=1}^{N}.
\label{eq:conventional_dataset}
\]
The computational cost of constructing
$\mathcal{D}_{\mathrm{solver}}$ is dominated by the repeated PDE solutions.
This cost becomes particularly significant for nonlinear, high-dimensional,
time-dependent, or geometrically varying equations.

The proposed physics-informed sampler reverses this procedure. It first
generates a candidate solution and then evaluates the governing operators
to construct the corresponding PDE inputs. The resulting input--solution
pairs are obtained without repeatedly solving the differential equation.

\subsection{Physics-informed sampler}
\label{sec:solution_space_generator}

{To implement this reversed procedure, let}
$u_{\bm{\theta}}(t,\boldsymbol{x})$
denote a {Bayesian} neural network (BNN) representation of a candidate solution field, where
$\bm{\theta}$ collects all weights and biases of the network. A probability
distribution is assigned to the network parameters,
$
\bm{\theta}
\sim
P(\bm{\theta})$.
Consequently, $u_{\bm{\theta}}$ becomes a random function or stochastic process. Repeated draws of
$\bm{\theta}$ generate samples from an implicit probability measure $\mu_u$ on
a function space of candidate solutions.

For a fully connected neural network with $L$ layers, the spatial-temporal
coordinate is first represented as
\begin{equation}
\boldsymbol{h}^{(0)}
=
\left(
t,
\boldsymbol{x}
\right).
\label{eq:bnn_input}
\end{equation}
The hidden layers are given by
\begin{equation}
\boldsymbol{h}^{(\ell)}
=
\sigma
\left(
W^{(\ell)}
\boldsymbol{h}^{(\ell-1)}
+
\boldsymbol{b}^{(\ell)}
\right),
\qquad
\ell=1,\ldots,L-1,
\label{eq:bnn_hidden_layers}
\end{equation}
and the output is
\begin{equation}
u_{\theta}(t,\boldsymbol{x})
=
W^{(L)}
\boldsymbol{h}^{(L-1)}
+
b^{(L)}.
\label{eq:bnn_output}
\end{equation}
The architecture, activation function, and parameter priors determine the
regularity, amplitude, and spectral characteristics of the generated
solution fields. For instance, independent zero-mean Gaussian priors are commonly used as the priors for the weights and
biases in each layer:
\begin{equation}
W_{ij}^{(\ell)}
\sim
\mathcal{N}
\left(
0,
\left(
\sigma_W^{(\ell)}
\right)^2
\right),
\qquad
b_i^{(\ell)}
\sim
\mathcal{N}
\left(
0,
\left(
\sigma_b^{(\ell)}
\right)^2
\right).
\label{eq:bnn_parameter_priors}
\end{equation}
{For time-dependent equations, the temporal and spatial directions may have}
different characteristic scales. We therefore permit anisotropic priors of
the form
\begin{equation}
W_{t,j}^{(1)}
\sim
\mathcal{N}
\left(
0,
\sigma_t^2
\right),
\qquad
W_{x_k,j}^{(1)}
\sim
\mathcal{N}
\left(
0,
\sigma_{x_k}^2
\right),
\label{eq:anisotropic_priors}
\end{equation}
where $\sigma_t$ controls temporal variability and
$\sigma_{x_k}$ controls variability along the $k$th spatial direction. {Importantly, the solution generator is not trained against a reference dataset.} Each
candidate solution is obtained directly through a random draw from the
prescribed parameter prior. In this sense, the Bayesian neural network
serves as a {\emph{training-free implicit generative model}} in the solution space.

{Once a candidate solution has been drawn, the PDE inputs follow directly from the governing operators.} For the $i$th realization of the network parameters, 
$
\theta^{(i)}
\sim
P(\bm{\theta}),
$
the generated solution field is
$
u^{(i)}(t,\boldsymbol{x})
=
u_{\bm{\theta}^{(i)}}(t,\boldsymbol{x}).
\label{eq:sampled_solution}
$
The corresponding forcing term is obtained by applying the governing
differential operator:
\begin{equation}
f^{(i)}(t,\boldsymbol{x})
=
\mathcal{N}_{\bm{\beta}^{(i)}}
\left[
u^{(i)}
\right]
(t,\boldsymbol{x}).
\label{eq:generated_forcing}
\end{equation}
Similarly, the boundary condition is generated as
\begin{equation}
b^{(i)}
(t,\boldsymbol{x}_{\mathrm{bc}})
=
\mathcal{B}_{\bm{\beta}^{(i)}}
\left[
u^{(i)}
\right]
(t,\boldsymbol{x}_{\mathrm{bc}}),
\label{eq:generated_boundary_condition}
\end{equation}
and the initial condition is obtained from
\begin{equation}
u_0^{(i)}(\boldsymbol{x})
=
u^{(i)}(0,\boldsymbol{x}).
\label{eq:generated_initial_condition}
\end{equation}
All spatial and temporal derivatives appearing in
$\mathcal{N}_{\bm{\beta}}$ and $\mathcal{B}_{\bm{\beta}}$ are evaluated through
automatic differentiation (AD), {as in B-PINNs} \cite{yang2021b}. Consequently, every generated tuple satisfies the governing equation automatically.
The resulting physics-informed dataset is
\[
\mathcal{D}_{\mathrm{PI}}
=
\left\{
\left(
f^{(i)},
b^{(i)},
u_0^{(i)},
\beta^{(i)},
\Omega_{\boldsymbol{x}}^{(i)};
u^{(i)}
\right)
\right\}_{i=1}^{N}.
\label{eq:pi_sampler_dataset}
\]

No numerical PDE solver is required to construct
$\mathcal{D}_{\mathrm{PI}}$. Each realization requires only a neural-network evaluation and AD. Since the samples are mutually
independent, the data-generation procedure is also embarrassingly parallel
and can be efficiently implemented on modern GPUs. {The complete procedure is summarized in}
Algorithm~\ref{alg:pi_sampler}.

\begin{algorithm}[t]
\caption{Physics-informed sampler}
\label{alg:pi_sampler}
\begin{algorithmic}[1]
\Require Number of samples $N$; parameter distribution $\mu_{\beta}$;
geometry distribution $\mu_{\Omega}$; mixture weights
$\{\pi_m\}_{m=1}^{M}$; BNN priors
$\{P^{(m)}\}_{m=1}^{M}$
\Ensure Physics-informed dataset $\mathcal{D}_{\mathrm{PI}}$
\State Initialize $\mathcal{D}_{\mathrm{PI}}\gets\varnothing$
\For{$i=1,\ldots,N$}
    \State Draw $\beta^{(i)}\sim\mu_{\beta}$
    \State Draw $\Omega_{\boldsymbol{x}}^{(i)}\sim\mu_{\Omega}$,
    when applicable
    \State Draw
    $m^{(i)}
    \sim
    \operatorname{Categorical}
    (\pi_1,\ldots,\pi_M)$
    \State Draw
    $\theta^{(i)}
    \sim
    p_{\Theta}^{(m^{(i)})}$
    \State Generate
    $u^{(i)}=u_{\theta^{(i)}}$
    \State Compute
    $f^{(i)}
    =
    \mathcal{N}_{\beta^{(i)}}[u^{(i)}]$
    using automatic differentiation
    \State Compute
    $b^{(i)}
    =
    \mathcal{B}_{\beta^{(i)}}[u^{(i)}]$
    \State Compute
    $u_0^{(i)}
    =
    \left.u^{(i)}\right|_{t=0}$,
    when applicable
    \State Add
    $\left(
    f^{(i)},
    b^{(i)},
    u_0^{(i)},
    \beta^{(i)},
    \Omega_{\boldsymbol{x}}^{(i)};
    u^{(i)}
    \right)$
    to $\mathcal{D}_{\mathrm{PI}}$
\EndFor
\State \Return $\mathcal{D}_{\mathrm{PI}}$
\end{algorithmic}
\end{algorithm}

\subsection{Bayesian neural network--Gaussian process correspondence}
\label{sec:bnn_gp_correspondence}

{Having defined the sampler computationally, we next interpret the statistical properties of its solution generator}
through the correspondence between Bayesian neural networks and Gaussian
processes. It is well known that a single-hidden-layer BNN with independently
sampled parameters converges to a Gaussian process as the hidden-layer width
approaches infinity, provided that the parameters are scaled appropriately \cite{neal2012bayesian,pearce2020expressive}.

In the present work, we employ Fourier-feature activations of the form
$\boldsymbol{\phi}(z)
=[\cos z,\sin z]^T$.
A representative single-hidden-layer random network can be expressed as
\begin{equation}
u_H(\boldsymbol{\xi})
=
\sqrt{\frac{1}{{H}}}
\sum_{j=1}^{H}
[a_j
\cos
\left(
\boldsymbol{\omega}_j^{\mathsf{T}}
\boldsymbol{\xi}
\right) + b_j
\sin
\left(
\boldsymbol{\omega}_j^{\mathsf{T}}
\boldsymbol{\xi}
\right)
],
\label{eq:random_fourier_bnn}
\end{equation}
where $H$ denotes the hidden-layer width, and $\bm{\xi}$ the input. The parameters are sampled from
\begin{equation}
a_j, b_j
\sim
\mathcal{N}
\left(
0,
\sigma_a^2
\right),
\qquad
\boldsymbol{\omega}_j
\sim
\mathcal{N}
\left(
\boldsymbol{0},
\Sigma_{\omega}
\right),
\label{eq:fourier_bnn_priors}
\end{equation}
In the infinite-width limit, the random network induces a stationary
Gaussian process with a covariance of the form \cite{neal2012bayesian,pearce2020expressive}
\begin{equation}
K_u
\left(
\boldsymbol{\xi},
\boldsymbol{\xi}'
\right)
=
{\sigma_a^2}
\exp
\left[
-\frac{1}{2}
\left(
\boldsymbol{\xi}
-
\boldsymbol{\xi}'
\right)^{\mathsf{T}}
\Sigma_{\omega}
\left(
\boldsymbol{\xi}
-
\boldsymbol{\xi}'
\right)
\right].
\label{eq:anisotropic_se_kernel}
\end{equation}
For an isotropic parameter prior,
\[
\Sigma_{\omega}
=
\ell^{-2}I,
\label{eq:isotropic_covariance}
\]
the covariance then reduces to the squared-exponential (SE) kernel
\begin{equation}
K_u
\left(
\boldsymbol{\xi},
\boldsymbol{\xi}'
\right)
=
{\sigma_a^2}
\exp
\left(
-\frac{
\left\|
\boldsymbol{\xi}
-
\boldsymbol{\xi}'
\right\|_2^2
}{
2\ell^2
}
\right).
\label{eq:isotropic_se_kernel}
\end{equation}
{In addition, for problems that have different characteristic frequencies in different dimensions, we can then employ anisotropic parameter priors as discussed in Eq. \eqref{eq:anisotropic_priors}. 
According to \cite{micchelli2006universal}, the SE kernel is a universal translation-invariant kernel, i.e., on any compact
set $\mathcal{K}\subset\mathbb{R}^{d}$, its reproducing kernel Hilbert space
is dense in $C(\mathcal{K})$ under the uniform norm. We use this property only as qualitative
guidance for choosing the Fourier-feature BNN employed in this work. It does
not constitute a universality guarantee for the finite-width BNN, nor does it
imply that a finite set of PI-sampler realizations covers all continuous
functions or matches an arbitrarily prescribed distribution.
}
The correlation length $\ell$ is inversely related to the scale of the
first-layer weight prior. Consequently, modifying the prior variance changes
the characteristic spatial and temporal scales of the generated functions.
{In the computations, we use finite-width BNNs rather than their infinite-width limits. Prior studies report that networks with more than 50 hidden units can already approximate their GP limits well in the settings examined \cite{pearce2020expressive,yang2020physics}; here, however, the correspondence is used only as guidance for prior design and interpretation.}



{This interpretation also suggests a simple way to broaden the sampled function class. A single Bayesian neural network prior may generate functions concentrated}
around a limited range of amplitudes or correlation lengths. To increase
the diversity of the generated solution fields, we employ a mixture of
parameter priors:
\[
P(\bm{\theta})
=
\sum_{m=1}^{M}
\pi_m
P^{(m)}(\bm{\theta}),
\label{eq:mixture_parameter_prior}
\]
where
\[
\pi_m
\geq
0,
\qquad
\sum_{m=1}^{M}
\pi_m
=
1.
\label{eq:mixture_weights}
\]
Sampling from the mixture consists of first selecting a component,
\[
m
\sim
\operatorname{Categorical}
\left(
\pi_1,
\ldots,
\pi_M
\right),
\label{eq:mixture_component_sampling}
\]
and then sampling the network parameters from the selected component:
$
\bm{\theta}
\sim
P^{(m)}.
\label{eq:component_parameter_sampling}
$
Each component uses different parameter variances and therefore generates
solution fields with different characteristic amplitudes, correlation
lengths, or spectral properties. The resulting measure over the solution
space is
\[
\mu_u
=
\sum_{m=1}^{M}
\pi_m
\mu_u^{(m)}.
\label{eq:mixture_solution_measure}
\]

For a fixed equation parameter $\bm{\beta}$, the corresponding forcing measure
is
\[
\mu_f
=
\left(
\mathcal{N}_{\bm{\beta}}
\right)_{\#}
\mu_u,
\label{eq:mixture_forcing_pushforward}
\]
which can equivalently be written as
\[
\mu_f
=
\sum_{m=1}^{M}
\pi_m
\left(
\mathcal{N}_{\bm{\beta}}
\right)_{\#}
\mu_u^{(m)}.
\label{eq:mixture_forcing_measure}
\]

For nonlinear differential operators, the individual pushforward
components  generally remain
analytically intractable. Nevertheless, each component can be sampled
directly. The mixture therefore provides a practical mechanism for
generating training data covering a wider range of functional scales. {The} mixture weights and prior variances are not selected to reproduce an
exact prescribed probability density for the forcing term. Instead, they
are chosen to generate sufficiently diverse candidate solutions for the
intended downstream operator-learning tasks. Their adequacy is evaluated
through empirical properties of the generated functions and through
prediction accuracy on independently generated testing data.

\subsection{Implicit pushforward distribution}
\label{sec:implicit_pushforward}

{The preceding construction admits a natural probabilistic interpretation as an implicit generative model in function space.} Let $X$ denote a function space for candidate solutions and $Y$ a function
space for forcing terms. For a fixed equation parameter $\bm{\beta}$, the
differential operator defines a mapping
\[
\mathcal{N}_{\beta}:
X
\longrightarrow
Y.
\label{eq:differential_operator_mapping}
\]
The BNN prior induces a probability measure $\mu_u$ on
$X$. Applying the differential operator to each realization produces the
pushforward measure
\[
\mu_f
=
\left(
\mathcal{N}_{\beta}
\right)_{\#}
\mu_u.
\label{eq:forcing_pushforward_measure}
\]
For any measurable subset $A\subseteq Y$, the pushforward measure is
defined by
\[
\mu_f(A)
=
\mu_u
\left(
\mathcal{N}_{\beta}^{-1}(A)
\right).
\label{eq:pushforward_definition}
\]
Thus, the distribution of the forcing term is not prescribed independently.
It is induced by passing the randomly generated solution field through the
governing differential operator.

When the equation parameter $\bm{\beta}$ is also random, we introduce the joint transformation
\[
\mathcal{T}:
\mathcal{P}\times X
\longrightarrow
Y,
\qquad
\mathcal{T}(\bm{\beta},u)
=
\mathcal{N}_{\bm{\beta}}[u].
\label{eq:joint_transformation}
\]
Assuming, for simplicity, that $\bm{\beta}$ and $u$ are sampled independently,
\[
(\bm{\beta},u)
\sim
\mu_{\beta}\otimes\mu_u,
\label{eq:joint_sampling_beta_u}
\]
the induced forcing measure is
\[
\mu_f
=
\mathcal{T}_{\#}
\left(
\mu_{\beta}\otimes\mu_u
\right).
\label{eq:joint_pushforward_measure}
\]

{We now discuss the induced forcing measure for linear and nonlinear PDE problems, respectively. For linear equations, the induced forcing distribution can be characterized analytically.} Suppose that the solution prior is represented by a Gaussian process,
\[
u
\sim
\mathcal{GP}
\left(
m_u,
K_u
\right),
\label{eq:solution_gp}
\]
and let $\mathcal{L}_{\beta}$ denote a linear differential operator. The
forcing term is then
$
f
=
\mathcal{L}_{\beta}[u].
\label{eq:linear_forcing}
$
Since a linear transformation of a Gaussian process remains Gaussian,
\[
f
\sim
\mathcal{GP}
\left(
m_f,
K_f
\right),
\label{eq:forcing_gp}
\]
where
\[
m_f(\boldsymbol{\xi})
=
\mathcal{L}_{\beta,\boldsymbol{\xi}}
m_u(\boldsymbol{\xi}), ~ K_f
\left(
\boldsymbol{\xi},
\boldsymbol{\xi}'
\right)
=
\mathcal{L}_{\beta,\boldsymbol{\xi}}
\mathcal{L}_{\beta,\boldsymbol{\xi}'}
K_u
\left(
\boldsymbol{\xi},
\boldsymbol{\xi}'
\right), ~ \boldsymbol{\xi}
=
\left(
t,
\boldsymbol{x}
\right).
\label{eq:forcing_gp_mean}
\]
and the subscripts on $\mathcal{L}_{\beta}$ indicate the argument of the
covariance function on which the differential operator acts. The above equations provide an analytical connection between
the prior in the solution space and the induced statistics in the forcing
space. They may therefore be used to guide the design of the solution prior
for linear equations. 


For a nonlinear differential operator, the induced measure $\mu_f$ is
generally non-Gaussian and does not admit a tractable probability density
or closed-form stochastic-process representation. Nonlinear terms involving
products, powers, or nonlinear functions of $u$ and its derivatives generally produce complicated dependencies and higher-order statistics.

Importantly, the proposed sampler does not require explicit evaluation of
the density of $\mu_f$ for implementation in both linear and nonlinear problems. Direct samples are obtained through
\[
\theta
\sim
P(\bm{\theta}),
\qquad
u
=
u_{\bm{\theta}},
\qquad
f
=
\mathcal{N}_{\bm{\beta}}[u].
\label{eq:implicit_sampling_process}
\]
This is analogous to an implicit generative model, in which samples can be
generated by a known stochastic transformation even when the corresponding
probability density cannot be evaluated analytically. The relevant
computational property is therefore the ability to sample from $\mu_f$,
rather than the availability of a closed-form expression for its density in the present PI-sampler.



{It is also worth mentioning that the pushforward formulation provides a qualitative interpretation of which forcing functions are not excluded by the sampler.} Let $X$ and $Y$ be Banach spaces chosen such that
\[
\mathcal{N}_{\beta}:
X
\longrightarrow
Y,
\label{eq:continuous_operator}
\]
is continuous. Consider a target solution $u^{\ast}\in X$ and its
corresponding forcing term
$
f^{\ast}
=
\mathcal{N}_{\beta}
\left[
u^{\ast}
\right].
$
If
$
u^{\ast}
\in
\operatorname{supp}(\mu_u),
$
then continuity of $\mathcal{N}_{\beta}$ implies
\[
f^{\ast}
\in
\operatorname{supp}
\left[
\left(
\mathcal{N}_{\beta}
\right)_{\#}
\mu_u
\right].
\label{eq:support_inclusion}
\]
Indeed, every open neighborhood of $u^{\ast}$ has positive probability
under $\mu_u$, and continuity maps a sufficiently small neighborhood of
$u^{\ast}$ into a neighborhood of $f^{\ast}$. Thus, the induced forcing
measure assigns positive probability to every neighborhood of $f^{\ast}$.

This observation should be interpreted qualitatively. Support membership
does not guarantee that a finite dataset contains a sample close to
$f^{\ast}$. A target forcing may belong to the support while the probability
mass in its neighborhood remains very small. Practical coverage therefore
depends on the concentration of the induced measure, the number of
generated samples, and the prior configuration. The effectiveness of the
sampler is ultimately assessed empirically rather than through support
membership alone.

{The homogeneous-forcing case illustrates this distinction.} The exact event
$
f
\equiv
0
$
typically has probability zero under a nondegenerate continuous stochastic
prior and is therefore almost surely absent from a finite randomly generated
dataset. Nevertheless, if the corresponding homogeneous solution lies in
the support of $\mu_u$, then forcing functions arbitrarily close to zero are
not excluded from the induced forcing distribution. It is empirically found that including low-frequency
 prior components may increase the probability of sampling
near this regime.

{The preceding construction yields five practical properties. First, the PI-sampler avoids}
repeated numerical PDE solutions during data generation. Each training pair
is obtained through random function generation and automatic
differentiation (AD). Second, the same procedure applies to both linear and nonlinear equations. For nonlinear equations, the induced forcing distribution is generally  implicit, but this does not prevent direct sampling.  Third, the method does not require analytically known eigenfunctions or a problem-specific finite basis as in \cite{hasani2025generating}. The stochastic neural network can be
instantiated for steady and time-dependent equations and for both low  and high spatial dimensions. Fourth, all realizations are independent and can therefore be generated in
parallel. {Fifth,} the physics-informed sampler is independent of the downstream
operator architecture. It produces paired functional data and can be used
to train DeepONet, Fourier neural operators, Transformer-based operators,
or other operator-learning models. In Sec.~\ref{sec:lto}, we introduce the
latent Transformer operator used in the present work to process the
heterogeneous and high-dimensional functional data generated by the
sampler.

\section{Latent Transformer operator}
\label{sec:lto}

The PI-sampler introduced in Sec.~\ref{sec:pi_sampler}
generates continuous realizations of the source term, boundary and initial
conditions, and solution fields. These functions can be evaluated at
arbitrary spatial and temporal coordinates. {However,} directly using
their pointwise values as inputs to an operator-learning model becomes
inefficient when the computational dimension is high, the geometry varies
between samples, or several heterogeneous functional inputs must be
processed simultaneously.

To address these challenges, we employ a latent Transformer operator
(LTO), which combines a function encoder with a Transformer.
The function encoder represents the continuous fields using a compact set
of coefficients associated with learned coordinate-dependent basis
functions. The Transformer subsequently learns the mapping from the latent
representations of the PDE inputs to that of the solution. The resulting
architecture follows the general encoder-operator-decoder paradigm used
in latent-space operator learning
{\cite{kontolati2024learning,serrano2023operator,
wang2024latent,bahmani2025resolution}}. We note that the purpose here is not to introduce latent operator learning as a new principle, but to construct a practical
downstream operator that can process the heterogeneous and
high-dimensional data generated by the PI-sampler.


The resulting pipeline is
\[
\bm{\Lambda}
\xrightarrow{\mathrm{encoder}}
\bm{\lambda}
\xrightarrow{\mathrm{operator}}
\widehat{\psi}_u
\xrightarrow{\mathrm{decoder}}
\widehat u,
\]
where \(\bm{\Lambda}\) denotes the physical inputs required by the particular PDE, and \(\bm{\lambda}\) collects their encoded representations. Depending on the problem, these inputs include forcing, boundary and initial data, equation parameters, and geometry. For second-order time-dependent equations, the initial data include both the solution and its first temporal derivative. Figure \ref{fig:model architecture} summarizes the pipeline, and \ref{sec:lto_details} provides the architectural and optimization details.

\subsection{Function encoder}

Let \(\gamma\) denote a functional quantity, such as the forcing \(f\) or solution \(u\), and let \(\bm{\xi}\) denote its spatial or space–time coordinates. The function encoder approximates \(\gamma\) through
\begin{equation}
\gamma(\bm{\xi}) \approx 
\gamma_{\mathrm{FE}}(\bm{\xi})
= 
\sum_{i=1}^{p}\psi_{\gamma,i}\phi_i(\bm{\xi})
=
\psi_\gamma^{T}\phi(\bm{\xi}),
\end{equation}
where \(\phi(\xi)=(\phi_1(\bm{\xi}),\ldots,\phi_p(\bm{\xi}))^{T}\) contains learned continuous basis functions and \(\psi_\gamma\in\mathbb R^p\) is the coefficient vector. A coefficient network maps sampled function values \(U_\gamma\) to \(\psi_\gamma\), while a coordinate network evaluates the basis:
\[
\psi_\gamma=\mathrm{NN}_{C,\gamma}(U_\gamma),
\qquad
\phi(\xi)=\mathrm{NN}_{B}(F(\bm{\xi})).
\]
Here, \(F\) denotes an optional coordinate-feature expansion. Separate coefficient networks are used for different functional quantities; the basis network may be shared when appropriate.

The FE is trained using PI-sampler realizations by minimizing the reconstruction loss
\begin{equation}
\mathcal L_{\mathrm{FE},\gamma}
=
\frac{1}{B_{\mathrm{FE}}N_{\bm{\xi}}}
\sum_{k=1}^{B_{\mathrm{FE}}}
\sum_{j=1}^{N_{\bm{\xi}}}
\left|
\gamma^{(k)}(\bm{\xi}_j)
-
\bigl(\psi_\gamma^{(k)}\bigr)^T\phi(\bm{\xi}_j)
\right|^2.
\end{equation}
The sampler provides continuous fields that can be evaluated on a common point set for efficient batch training. The separable reconstruction resembles the branch–trunk decomposition of DeepONet \cite{lu2021learning}; here, it is used to learn functional representations before training the input–solution mapping.

At inference, inputs sampled on the training discretization are encoded directly by the coefficient network. For observations on a different or irregular point set, the basis is retained and the coefficients are recovered by optimization, as detailed in \ref{sec:lto_details}. This accommodates changing observation locations without retraining the basis, although recovery introduces additional computation and depends on the observation coverage and basis conditioning.

\subsection{Latent operator learning}

The functional inputs are encoded as coefficient vectors or represented by prescribed boundary and initial observations. Equation parameters and geometry are represented using the quantities appropriate to each problem. These representations are projected into a common embedding space and supplied as tokens to the Transformer. Type embeddings distinguish the input modalities, and coefficient vectors may be divided into multiple tokens when needed.  The Transformer learns the latent mapping
\[
\mathcal G_{\mathrm{LTO}}:
\bm{\lambda}\longmapsto\widehat{\psi}_u.
\]
The predicted solution is reconstructed using the pretrained solution basis:
\begin{equation}
\widehat u(\bm{\xi})
=
\sum_{i=1}^{p}\widehat{\psi}_{u,i}\phi_i(\bm{\xi}).
\end{equation}
The continuous basis permits evaluation at arbitrary query coordinates, with accuracy determined by the learned representation and predicted coefficients.

After FE training, the encoder and basis are fixed, and the Transformer is trained against the reference solution coefficients:
\begin{equation}
\mathcal L_{\mathrm{OL}}
=
\frac{1}{B_{\mathrm{OL}}}
\sum_{k=1}^{B_{\mathrm{OL}}}
\left\|
\widehat{\psi}_u^{(k)}
-
\psi_u^{(k)}
\right\|_2^2.
\end{equation}
Thus, operator learning takes place in the learned coefficient space. Token construction, Transformer blocks, and an optional field-space loss are described in \ref{sec:lto_details}. 

\subsection{Optional physics-informed refinement}

A pretrained LTO prediction may be refined for a specific test instance by optimizing only its {\emph solution coefficients}, following the physics-informed neural networks \cite{raissi2019physics}. Starting from prediction from the Transformer \(\psi=\widehat{\psi}_u\), we define
\[
u_\psi(\xi)=\sum_{i=1}^{p}\psi_i\phi_i(\bm{\xi})
\]
and minimize the following physics-informed loss
\begin{equation}
\mathcal L_{\mathrm{FT}}
=
\lambda_{\mathrm{eq}}\mathcal L_{\mathrm{eq}}
+
\lambda_{\mathrm{bc}}\mathcal L_{\mathrm{bc}}
+
\lambda_{\mathrm{ic}}\mathcal L_{\mathrm{ic}}.
\end{equation}
The terms measure the residual of the governing equation, discrepancies in the boundary conditions, and initial data, respectively. In addition, $\lambda_{eq}/\lambda_{bc}/\lambda_{ic}$  denote the weights for each loss term. For second-order time-dependent equations, the initial-data loss includes both the solution evaluated at the initial time as well as its temporal derivative. Conditions imposed exactly through the representation do not require an additional penalty.

During refinement, the basis and Transformer parameters remain fixed. The optimization therefore involves only the \(p\) solution coefficients, with the LTO prediction providing their initialization. We distinguish direct, zero-shot predictions from refined predictions throughout the numerical results. The individual loss terms are defined in \ref{sec:lto_details}.

\begin{figure}[H]
    \centering    \includegraphics[width=\textwidth]{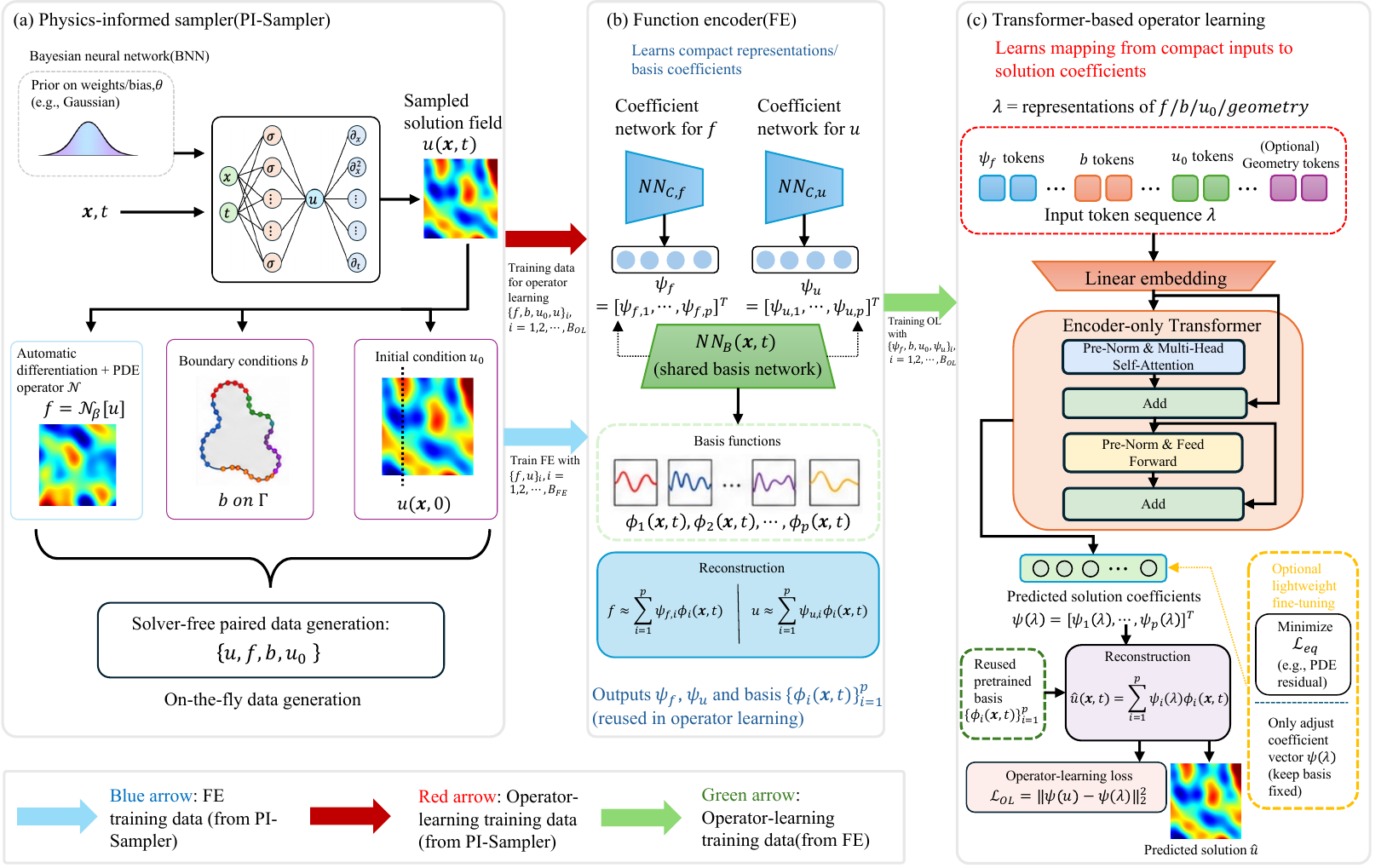}
    \caption{\label{fig:model architecture}
    Schematic of PI-sampler and latent Transformer operator, which consists of three parts: (a) Physics-informed sampler (PI-sampler) for generating training data, (b) Function encoder (FE) for learning compact representations of $u$ and $f$, and (c) operator learning using a Transformer. The blue arrow represents the FE training data generated from PI-sampler: $\{f,u\}_i, i = 1, ..., B_{FE}$ ($B_{FE}$ denotes the batch size),  Red arrow represents the training data for operator learning generated from PI-sampler:$\{f,b, u_0, u\}_j, j= 1, ..., B_{OL}$ ($B_{OL}$ denotes the batch size), $\bm{\lambda}$ denotes the representations of $f/b/u_0$, and possibly the geometries. 
    }
\end{figure}

{Together, the components form a sequential pipeline as shown in Fig. \ref{fig:model architecture}:} the PI-sampler generates PDE-consistent training data, the
function encoder constructs compact and continuously evaluable functional
representations, the Transformer learns the mapping between heterogeneous
latent inputs and solution coefficients, and the optional refinement
corrects the predicted coefficients using the governing equation.

\section{Results and discussion}
\label{sec:results}

In this section, we evaluate the proposed framework on a series of
increasingly challenging problems, including a one-dimensional nonlinear
reaction-diffusion equation, a Poisson equation on parameterized
geometries, a time-dependent Sine-Gordon equation, a five-dimensional
nonstationary PDE, and vortex-induced vibration of a flexible cylinder.
Unless otherwise stated, the LTO predictions obtained directly from the
pretrained model are referred to as zero-shot predictions. In cases where
the latent solution coefficients are subsequently optimized using the
physics-informed loss in Sec.~\ref{sec:lto}, the results are
referred to as refined predictions.

The relative $L_2$ error is employed as the primary accuracy metric:
\[
E
=
\frac{
\left\|
u_{\mathrm{pred}}
-
u_{\mathrm{ref}}
\right\|_2
}{
\left\|
u_{\mathrm{ref}}
\right\|_2
},
\label{eq:relative_l2_error}
\]
where $u_{\mathrm{pred}}$ and $u_{\mathrm{ref}}$ denote the predicted and
reference solutions, respectively. The reported errors are averaged over
all test samples unless otherwise specified. Details of the BNN priors,
training datasets, numerical reference solvers, and network configurations
in each case are provided in {\ref{app:data_generation} and
\ref{app:implementation_details}}. In all test cases, the training data are generated on the fly where required in the training of LTO, unless otherwise stated. Example codes are available at https://github.com/UQ4MI/PI-Sampler-NO.

\subsection{Nonlinear reaction--diffusion equation}
\label{sec:reaction_diffusion}

We first consider the following one-dimensional nonlinear
reaction--diffusion equation:
\begin{equation}
D
\frac{\partial^2 u}{\partial x^2}
-
u^3
=
f,
\qquad
x\in[-1,1],
\label{eq:reaction_diffusion}
\end{equation}
where $D=0.1$ is the diffusion coefficient. We initially impose Dirichlet
boundary conditions,
\begin{equation}
u(-1)
=
a,
\qquad
u(1)
=
b.
\label{eq:reaction_diffusion_dirichlet}
\end{equation}
The objective is to learn the solution operator
\[
\mathcal{G}_{\mathrm{RD}}:
(f,a,b)
\longmapsto
u.
\label{eq:reaction_diffusion_operator}
\]

The PI-sampler generates candidate solutions $u$ from multiple BNN priors
and evaluates Eq.~\eqref{eq:reaction_diffusion} through automatic
differentiation to obtain the corresponding forcing functions. The
boundary values are obtained by evaluating each candidate solution at
$x=-1$ and $x=1$.


We consider three types of test data after training the LTO.
First, in-distribution test samples are generated by drawing new parameter
realizations from the same BNN priors employed to construct the training
dataset. Specifically, 200 solution samples are drawn from each BNN prior,
and the corresponding forcing terms and boundary values are generated using
the PI-sampler. 

Second, to assess generalization to independently prescribed forcing
distributions, we draw forcing functions from GPs with
SE kernels:
\[
f
\sim
\mathcal{GP}
\left(
0,
K_l
\right),
\label{eq:reaction_diffusion_gp_test}
\]
where
\[
K_l(x,x')
=
\exp
\left[
-\frac{
(x-x')^2
}{
2l^2
}
\right], ~ l
\in
\left\{
0.5,
0.3,
0.1,
0.05
\right\}.
\label{eq:reaction_diffusion_gp_kernel}
\]
For each GP with different correlation length $l$,  1000 forcing realizations are considered. Further, the
corresponding boundary conditions are generated from the same BNN families
used during training to pair with the forcing realizations, and the reference solutions are computed using a
Runge-Kutta solver in MATLAB.

Third, we consider the homogeneous forcing case,
$
f
\equiv
0.
$
The exact zero function has probability zero under the continuous random
priors used by the PI-sampler and is therefore not expected to occur
exactly in a finite training dataset. The boundary conditions are generated in the same way as in the second scenario. This test examines whether training
on PI-sampler realizations enables the learned operator to generalize to the homogeneous regime.

\begin{figure}[H]
    \centering
    \includegraphics[width=\textwidth]{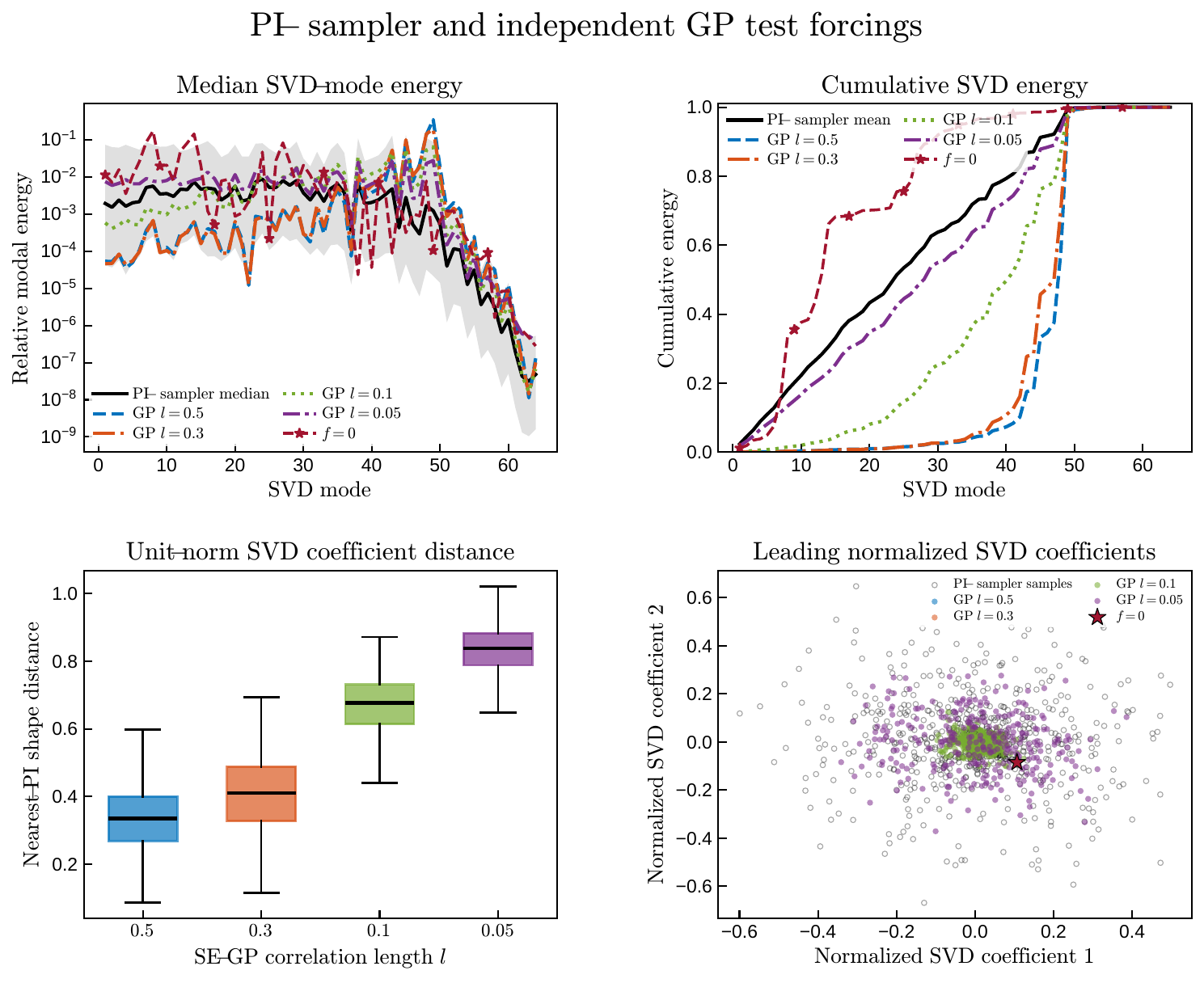}
    \caption{{
    Empirical SVD comparison of the centered forcing fields considered in the
    reaction--diffusion experiment. All ensembles are represented by
    $f-\overline{f}$ in a common SVD basis, where $\overline{f}$ is the
    empirical mean used for centering. The panels compare the modal-energy
    profiles, cumulative modal energy, nearest-PI-sampler distance after
    unit-norm projection, and the first two normalized SVD coefficients.
    The marker labeled $f=0$ therefore represents $-\overline{f}$ in the
    centered coordinates rather than a nonzero raw forcing.}}
    \label{fig:reaction_diffusion_svd_audit}
\end{figure}


{
{
Before presenting the prediction errors, we examine how the prescribed test
forcings relate to the empirical distribution induced by the BNN prior.
Figure~\ref{fig:reaction_diffusion_svd_audit} compares the PI-sampler and test
ensembles after centering and projecting them onto a common singular value
decomposition (SVD) basis computed from the PI-sampler forcing samples. The
modal-energy profiles, cumulative energy, and overlap of the leading normalized
SVD coefficients show that the test families, including the centered
representation of $f=0$, occupy regions represented by the PI-sampler ensemble
in this finite-dimensional projection. However, the nearest-sample distance
increases as the GP correlation length decreases from $l=0.5$ to $l=0.05$,
indicating that short-correlation-length forcings lie in progressively less
dense regions of the empirical distribution. Thus, the analysis supports the
relevance of the PI-sampler distribution to the particular test families used
below, but it does not establish coverage of arbitrary forcing functions or
distributions. Because the analysis is performed on $f-\overline{f}$, the
nonzero curve and marker associated with $f=0$ represent the centered field
$-\overline{f}$ rather than the raw zero forcing.
}
}


\begin{figure}[H]
    \centering

    \begin{minipage}[b]{0.31\linewidth}
        \centering
        \begin{overpic}[width=\linewidth]{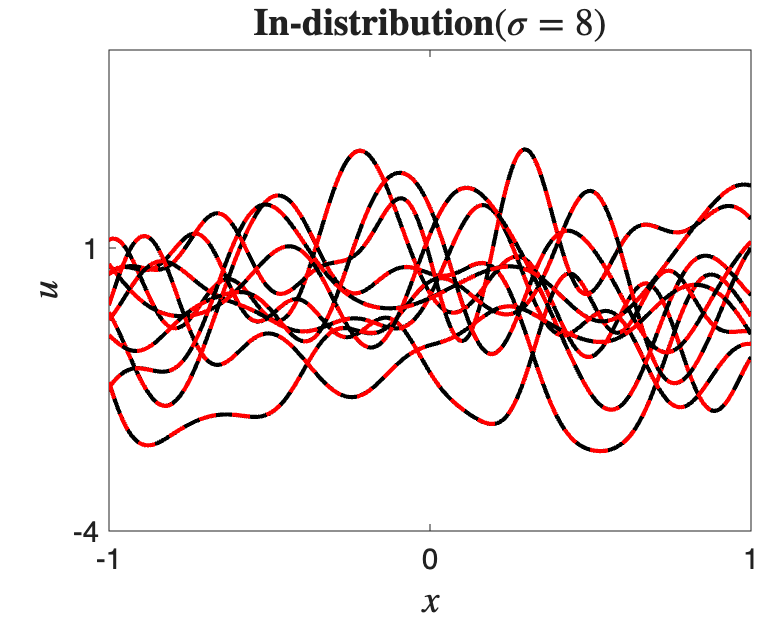}
            \put(4,80){\colorbox{white}{\makebox[1.2em][l]{\bfseries a}}}
        \end{overpic}
    \end{minipage}
    \hfill
    \begin{minipage}[b]{0.31\linewidth}
        \centering
        \includegraphics[width=\linewidth]{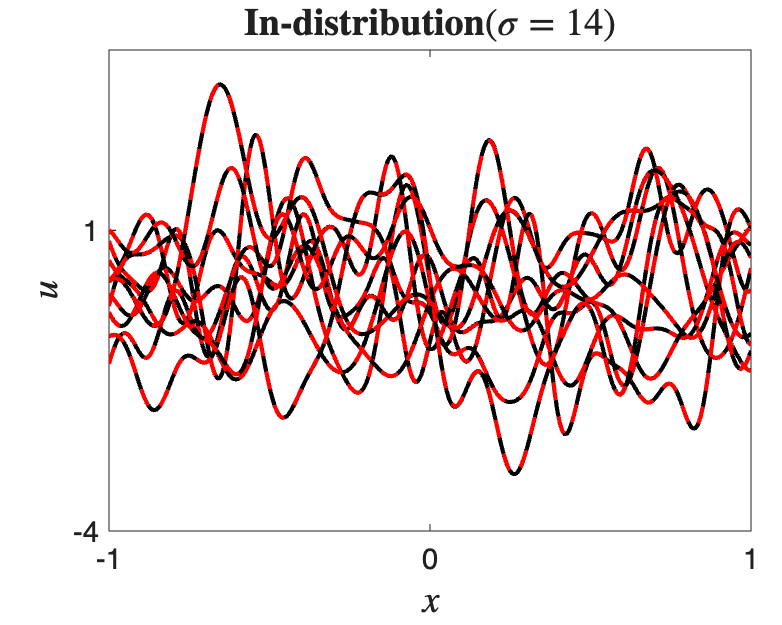}
    \end{minipage}
    \hfill
    \begin{minipage}[b]{0.31\linewidth}
        \centering
        \includegraphics[width=\linewidth]{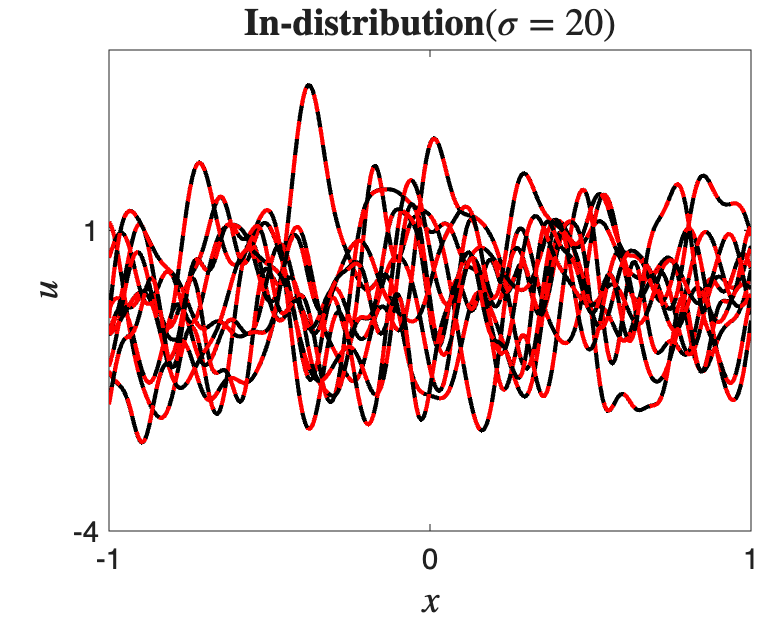}
    \end{minipage}

    \vspace{0.5em}

    \begin{minipage}[b]{0.31\linewidth}
        \centering
        \begin{overpic}[width=\linewidth]{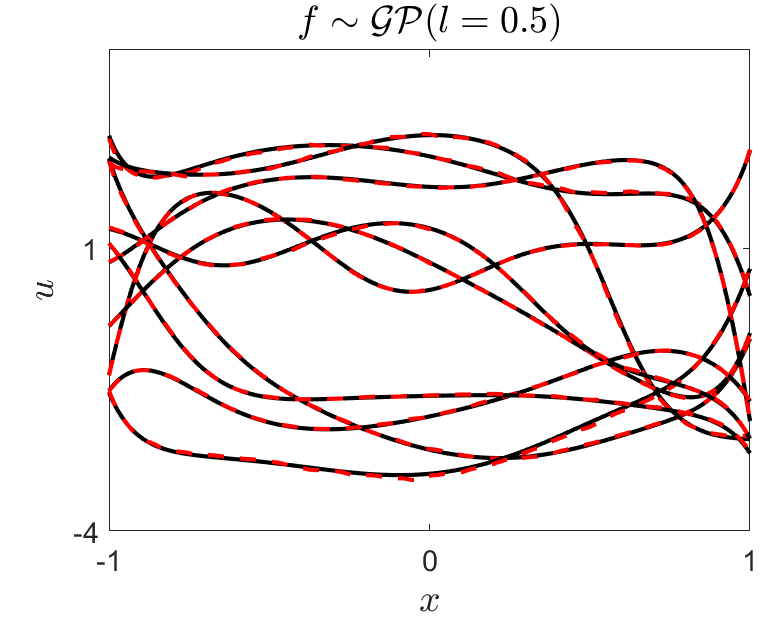}
            \put(4,80){\colorbox{white}{\makebox[1.2em][l]{\bfseries b}}}
        \end{overpic}
    \end{minipage}
    \hfill
    \begin{minipage}[b]{0.31\linewidth}
        \centering
        \includegraphics[width=\linewidth]{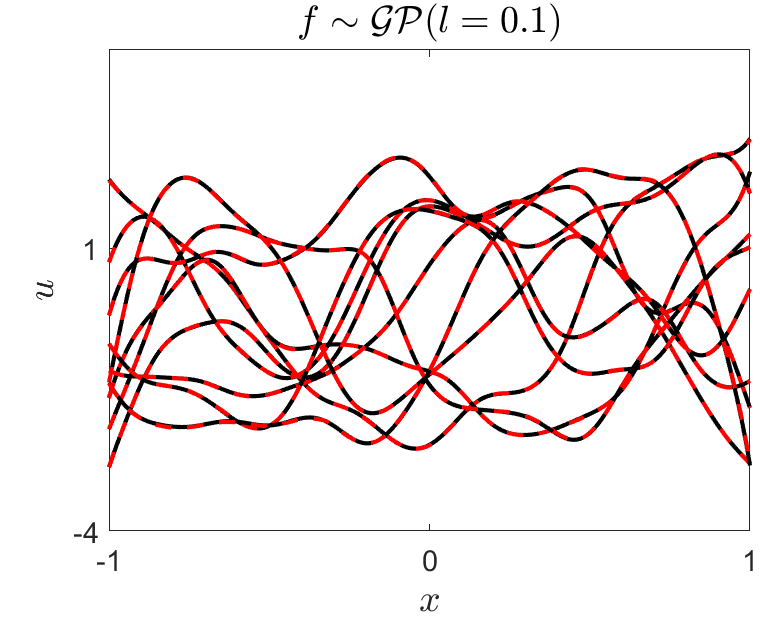}
    \end{minipage}
    \hfill
    \begin{minipage}[b]{0.31\linewidth}
        \centering
        \includegraphics[width=\linewidth]{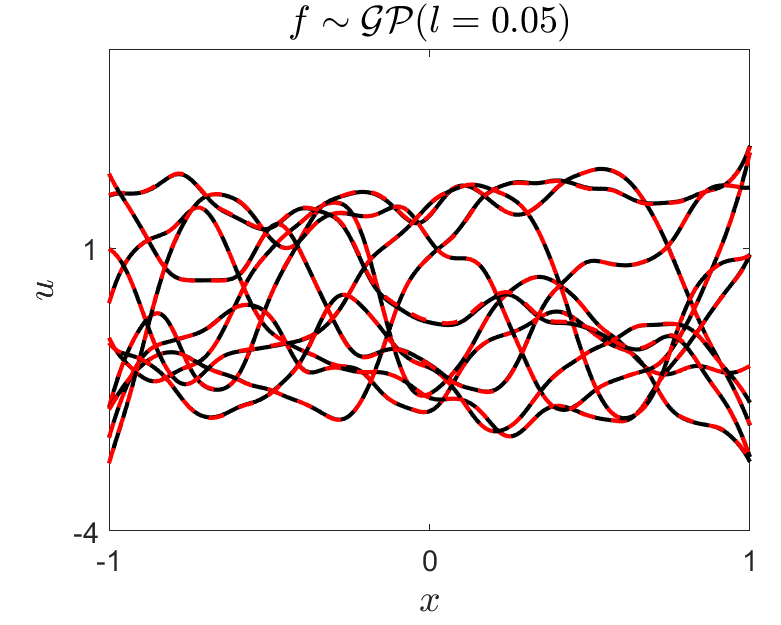}
    \end{minipage}

    \vspace{0.5em}

    \begin{minipage}[b]{0.31\linewidth}
        \centering
        \begin{overpic}[width=\linewidth]{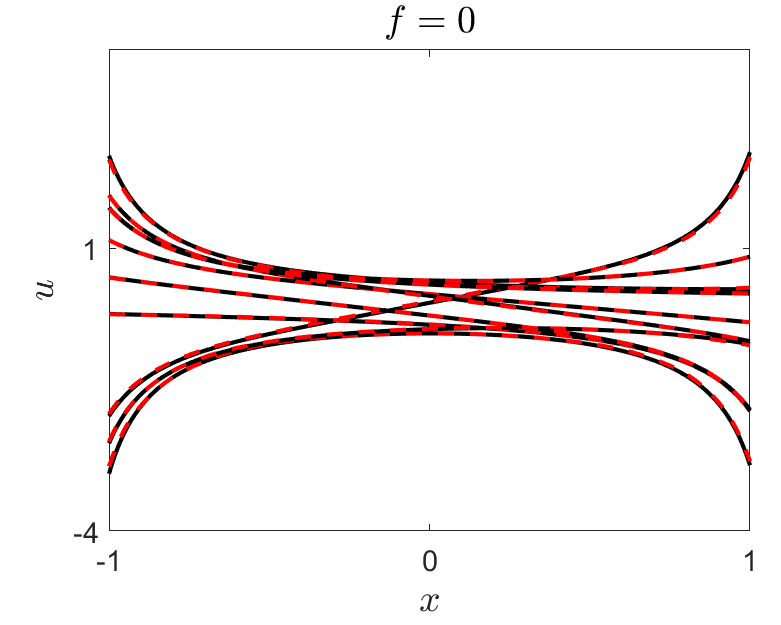}
            \put(4,80){\colorbox{white}{\makebox[1.2em][l]{\bfseries c}}}
        \end{overpic}
    \end{minipage}
    \hfill
    \begin{minipage}[b]{0.31\linewidth}
        \centering
        \begin{overpic}[width=\linewidth]{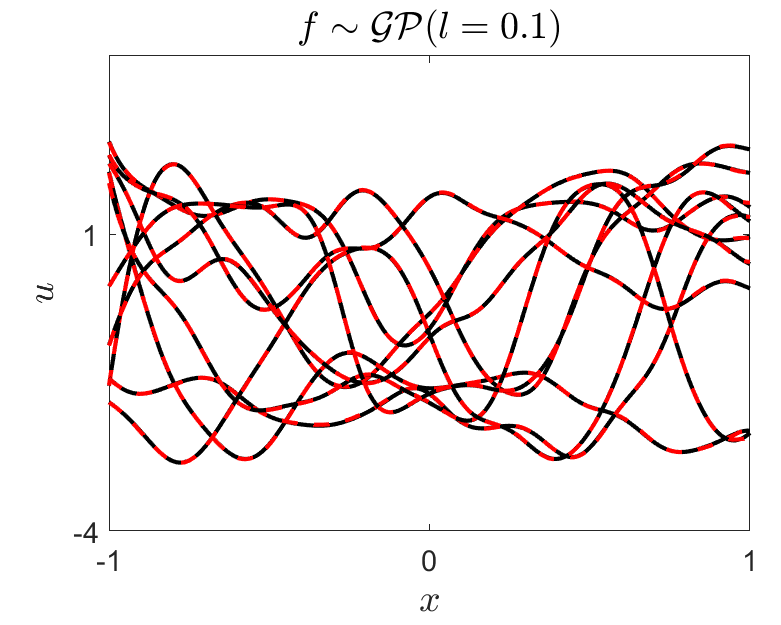}
            \put(4,80){\colorbox{white}{\makebox[1.2em][l]{\bfseries d}}}
        \end{overpic}
    \end{minipage}
    \hfill
    \begin{minipage}[b]{0.31\linewidth}
        \centering
        \includegraphics[width=\linewidth]{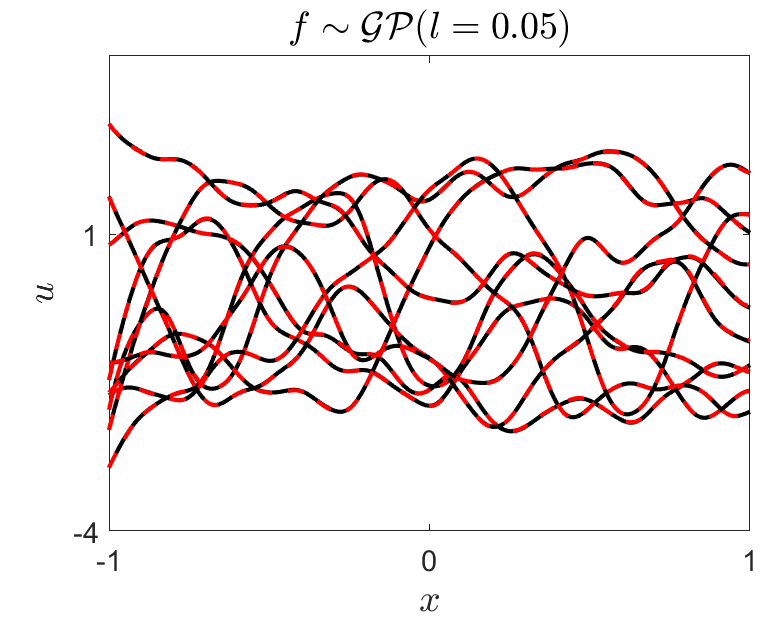}
    \end{minipage}

    \caption{\label{fig:section1-combined}
    LTO for steady nonlinear reaction-diffusion equation. Black solid lines denote the reference solutions, and red dashed lines denote the corresponding zero-shot predictions of LTO. Panels (a)--(c) show results under Dirichlet boundary conditions with prescribed values at both boundaries: (a) in-distribution test cases, (b)  cases with $f \sim \mathcal{GP}$, and (c) the special case $f=0$. Panel (d) shows results for mixed boundary conditions with $f \sim \mathcal{GP}$.} 
\end{figure}

{We now turn from empirical distributional overlap to predictive accuracy.} Representatives for the predictions from LTO in different cases and the mean relative $L_2$ errors are illustrated in Fig. \ref{fig:section1-combined} and
Table~\ref{tab:reaction_diffusion_dirichlet}, respectively.  The error is $0.47\%$ for
the in-distribution test set and $0.50\%$ for the forcing functions drawn
from the independent Gaussian-process distributions. The error increases
moderately to $1.37\%$ for $f=0$, but remains at the percent level.

{
To verify that the generated data are not tied to the LTO, we also train a
multi-input DeepONet on the same PI-sampler dataset. Both its branch and trunk
components are fully connected networks: we use three  branch networks to separately take as input the sourcing term, and the two boundary conditions, and each network has five hidden
layers of width 256 with ReLU activations. Further, one trunk network is employed, which has four
hidden layers of width 256 with hyperbolic-tangent activations. Both networks
produce 256-dimensional outputs. As shown in
Table~\ref{tab:reaction_diffusion_dirichlet}, the resulting DeepONet also
achieves percent-level errors on all three test sets, although its errors are
higher than those of the LTO for the first two categories. We do not interpret
this comparison as an architecture ranking because the DeepONet hyperparameters
were not exhaustively tuned; its purpose is to demonstrate that PI-sampler data
can train a different neural-operator architecture.
}

\begin{table}[t]
    \centering
    \caption{
    {Mean relative $L_2$ errors of the LTO and DeepONet predictions for the}
    nonlinear reaction--diffusion equation with Dirichlet boundary
    conditions.
    }
    \label{tab:reaction_diffusion_dirichlet}
    \begin{tabular}{lccc}
        \hline
        $E$
        &
        In-distribution
        &
        $f\sim\mathcal{GP}_{l \in \{0.5, 0.3, 0.1, 0.05 \}}$
        &
        $f=0$
        \\
        \hline
        LTO
        &
        $0.47\%$
        &
        $0.50\%$
        &
        $1.37\%$
        \\
         \hline
        DeepONet
        &
        $3.08\%$
        &
        $2.08\%$
        &
        $1.50\%$
        \\
        \hline
    \end{tabular}
\end{table}



{The preceding tests use Dirichlet conditions at both endpoints. To examine a derivative-dependent boundary operator, we retain the governing equation in
Eq.~\eqref{eq:reaction_diffusion} but replace the condition at}
$x=1$ by a Robin boundary condition:
\begin{equation}
u(-1)
=
a,
\qquad
u(1)
+
\frac{\partial u}{\partial x}(1)
=
b.
\label{eq:reaction_diffusion_mixed_bc}
\end{equation}
Similar as in the previous cases, three test categories are considered: in-distribution samples,
forcing terms drawn from the independent Gaussian processes, and the homogeneous forcing case. Further, the boundary conditions at the testing stage are also generated using the same procedure as in the preceding Dirichlet tests in the above test cases. 

We illustrate the representative predictions from LTO and the mean relative errors are in Fig. \ref{fig:section1-combined} and
Table~\ref{tab:reaction_diffusion_mixed}, respectively. The errors remain close to or
below $1\%$ for all test categories. These results demonstrate that the
PI-sampler can generate training data for boundary operators involving
both function values and derivatives without changing the underlying
sampling procedure.

\begin{table}[t]
    \centering
    \caption{
    Mean relative $L_2$ errors of the zero-shot LTO predictions for the
    nonlinear reaction--diffusion equation with the mixed boundary
    condition in Eq.~\eqref{eq:reaction_diffusion_mixed_bc}.
    }
    \label{tab:reaction_diffusion_mixed}
    \begin{tabular}{lccc}
        \hline
        Test set
        &
        In-distribution
        &
        $f\sim\mathcal{GP}_{l \in \{0.5, 0.3, 0.1, 0.05 \}}$
        &
        $f=0$
        \\
        \hline
        $E$
        &
        $0.45\%$
        &
        $0.45\%$
        &
        $1.21\%$
        \\
        \hline
    \end{tabular}
\end{table}

\subsection{Poisson equation on parameterized geometries}
\label{sec:poisson_geometry}

{Having examined variations in forcing and boundary conditions on a fixed interval, we next introduce geometric variability through the Poisson equation on two-dimensional domains:}
\begin{equation}
-k\Delta u(\boldsymbol{x})
=
f(\boldsymbol{x}),
\qquad
\boldsymbol{x}
\in
\Omega_{\boldsymbol{x}},
\label{eq:poisson_geometry}
\end{equation}
where $k=0.01$ is the constant diffusion coefficient. The equation is subject to the Dirichlet boundary condition 
\begin{equation}
u(\boldsymbol{x}_{\mathrm{bc}})
=
b(\boldsymbol{x}_{\mathrm{bc}}),
\qquad
\boldsymbol{x}_{\mathrm{bc}}
\in
\partial\Omega_{\boldsymbol{x}}.
\label{eq:poisson_geometry_bc}
\end{equation}
The computational domain varies among problem instances. The corresponding
operator is
\[
\mathcal{G}_{\mathrm{P}}:
\left(
f,
b,
\Omega_{\boldsymbol{x}}
\right)
\longmapsto
u.
\label{eq:poisson_geometry_operator}
\]


Specifically, the domain boundary is represented in polar coordinates by a radius function $r(\alpha)$, where
$
\alpha
\in
[0,2\pi].
$
To ensure closure of the curve, the angular coordinate is warped according to
\begin{equation}
\alpha
\longmapsto
\left[
\cos(\alpha),
\sin(\alpha)
\right].
\label{eq:geometry_input_warping}
\end{equation}
A separate BNN maps the warped angular coordinate to an unconstrained
radius output $r_{\mathrm{BNN}}(\alpha)$. The physical radius is then
defined by
\begin{equation}
r(\alpha)
=
0.45
\tanh
\left[
0.3
r_{\mathrm{BNN}}(\alpha)
\right]
+
0.55.
\label{eq:geometry_radius_constraint}
\end{equation}
This transformation restricts the generated radius to a prescribed range
and produces {diverse closed domains.}

\subsection{Performance of LTO trained on data generated from PI-sampler}

Following the coordinate-transformation strategy introduced in {\cite{xiao2024deformation}}, a homeomorphic mapping
\[
\mathcal{D}
\left(
\partial\Omega_{\boldsymbol{x}}
\right):
\Omega_{\boldsymbol{x}}
\longrightarrow
\Omega_0,
\label{eq:geometry_homeomorphism}
\]
maps each physical domain to a reference unit disk $\Omega_0$. The LTO is trained in the reference domain. Predictions are mapped
back to the physical domain through
\[
\boldsymbol{x}
=
\mathcal{D}^{-1}
\left(
\partial\Omega_{\boldsymbol{x}}
\right)
(\boldsymbol{x}_0),
\qquad
\boldsymbol{x}_0
\in
\Omega_0.
\label{eq:inverse_geometry_mapping}
\]


{After defining the geometry representation, we assess the trained LTO using forcing functions drawn independently from isotropic Gaussian processes}
with correlation lengths
\[
l_x
=
l_y
\in
\left\{
0.15,
0.2,
0.3,
0.5
\right\},
\label{eq:poisson_test_lengths}
\]
together with the additional constant forcing cases
\[
f
=
1
\qquad
\text{and}
\qquad
f
=
0.
\label{eq:poisson_constant_forcings}
\]
The boundary conditions are either generated from GP realizations of $u$
with correlation lengths
$l\in \left\{ 0.15, 0.2, 0.3, 0.5 \right\},$ or set to $ b = 0.$

\begin{figure}[t]
   \centering
   \includegraphics[width=\textwidth]
   {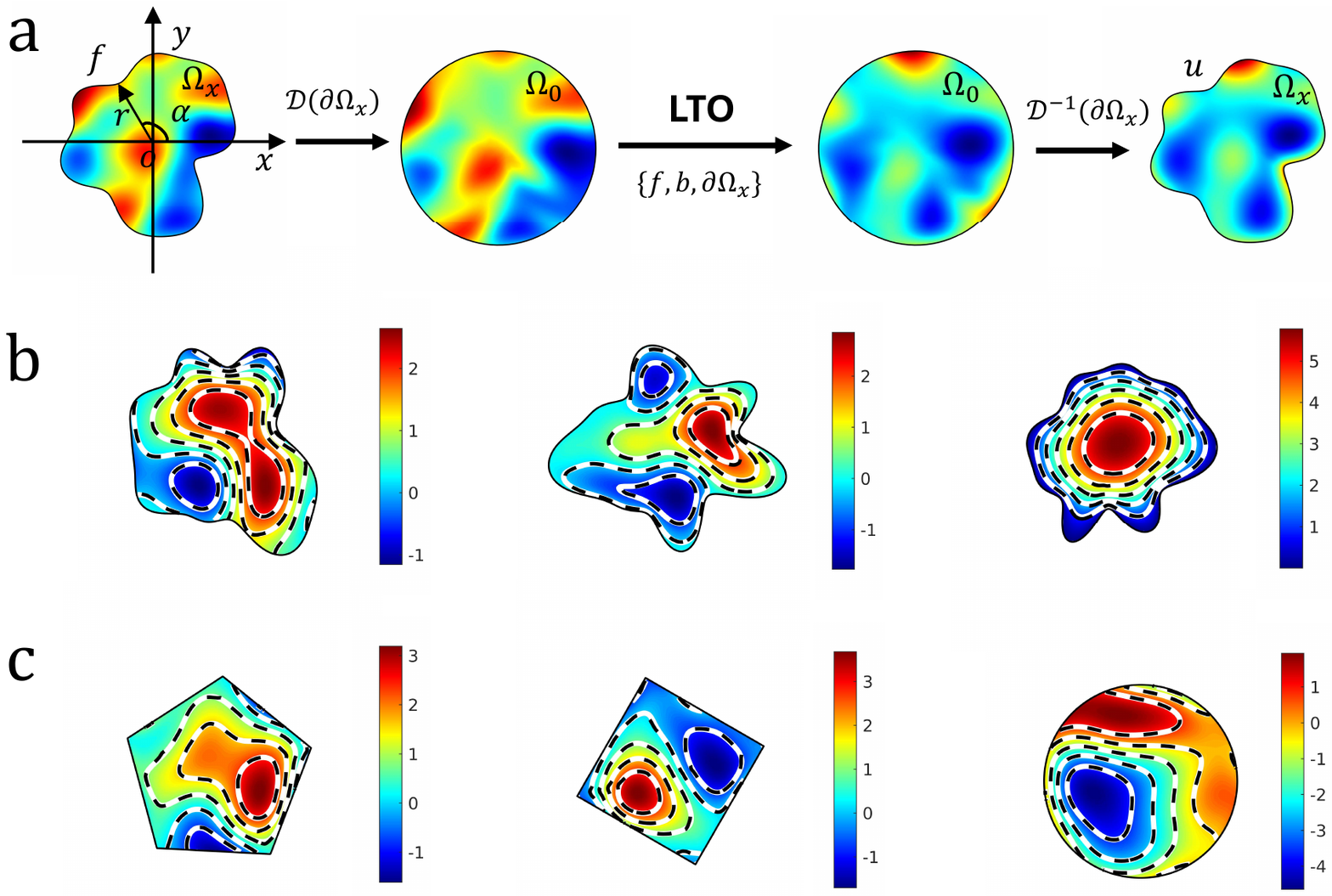}
   \caption{
   LTO predictions for the Poisson equation on parameterized geometries. Colored background and white solid lines denote the reference solutions, and the black dashed line represent the LTO predictions. 
   Panel (a) illustrates the mapping between the physical and reference
   domains. Panel (b) presents representative predictions for
   $(f\sim\mathcal{GP},b\sim\mathcal{GP})$,
   $(f\sim\mathcal{GP},b=0)$, and $(f=1,b=0)$.
   Panel (c) shows representative predictions on geometries not generated
   by the BNN geometry prior during training.
   }
   \label{fig:poisson_geometry_results}
\end{figure}


{The six forcing categories and five boundary-condition categories yield 29 test configurations after excluding the case with both zero forcing and zero boundary data. We additionally consider four configurations in which the forcing functions are obtained by applying the Poisson operator to new candidate solution fields sampled independently from the BNNs with isotropic priors where $\sigma^1_{W}  = 1/l,\,l\in \left\{ 0.15, 0.2, 0.3, 0.5 \right\}$. The additional configurations are labeled ``$f$(BNN)'' bringing the total number of configurations in Table 3 to 33.}

For each GP-based
configuration, 1000 realizations are generated. Reference solutions are
computed using the finite-element solver with automatically generated meshes in the MATLAB PDE Toolbox. We show several representative predictions from LTO in Fig. \ref{fig:poisson_geometry_results}, and it is observed that the results from the LTO are in good agreement with the reference solutions. Further, the mean relative errors for all combinations are summarized in
Table~\ref{tab:poisson_geometry_errors}. The errors remain below $3\%$
across all configurations. The results indicate that the data generated by
the PI-sampler support generalization to independently sampled GP forcings,
constant forcings, and multiple boundary-condition distributions. In each test case, the computational domain is  drawn from the BNN for generating geometries mentioned above. 

\begin{table}[t]
\centering
\caption{
Mean relative $L_2$ errors of the zero-shot LTO predictions for the
Poisson equation on parameterized geometries.
}
\label{tab:poisson_geometry_errors}
\begin{tabular}{c|ccccccc}
\toprule
\diagbox{$\bm{b}$}{$\bm{f}$} &$\mathcal{GP}_{l = 0.15}$ & $\mathcal{GP}_{l=0.2}$ & $\mathcal{GP}_{l=0.3}$  & $\mathcal{GP}_{l=0.5}$  & $f=1$ & $f=0$ & $f$ (BNN)  \\
\midrule
$\mathcal{GP}_{l = 0.15}$  & 1.74\% & 1.85\% & 2.41\% & 2.93\% & 2.66\% & 2.31\% & 1.47\% \\ 
$\mathcal{GP}_{l = 0.2}$   & 1.52\% & 1.68\% & 2.09\% & 2.65\% & 2.62\% & 1.44\% & 0.73\% \\ 
$\mathcal{GP}_{l = 0.3}$   & 1.37\% & 1.58\% & 1.87\% & 2.74\% & 2.57\% & 0.79\% & 0.45\% \\ 
$\mathcal{GP}_{l = 0.5}$  & 1.33\% & 1.58\% & 2.01\% & 2.41\% & 2.64\% & 0.51\% & 0.39\% \\ 
$\bm{b}=0$        & 1.48\% & 1.63\% & 2.12\% & 2.63\% & 2.43\% & \texttt{-} & \texttt{-} \\
\bottomrule
\end{tabular}
\end{table}

{
Beyond the uniformly low error levels, Table~\ref{tab:poisson_geometry_errors}
shows a modest increase in error for smoother forcing functions with larger
correlation lengths in several boundary-condition categories. This trend
reflects the joint influence of the forcing, boundary condition, geometry, and
learned latent representation; it should not be interpreted through a single
distance from the training distribution. Because the geometry-conditioned
pushforward distribution is not available explicitly, generalization is
assessed directly on these independently generated test families.
}


{Having varied the forcing and boundary distributions, we next examine whether the LTO generalizes to geometries that}
are not generated by the BNN geometry prior used during training. Three
representative domains are considered: a circle, a quadrilateral, and a
pentagon. These geometries share the same topology as the training domains
but exhibit qualitatively different boundary shapes.
For each geometry, 1000 forcing functions are drawn from
$
f
\sim
\mathcal{GP}_{l_x=l_y = 0.2},
$
and the boundary conditions are generated from independent realizations of
the same Gaussian process.

The results are summarized in
Table~\ref{tab:unseen_geometry_errors}. The errors are below $3\%$ for
all three domains, indicating that the reference-domain representation
provides useful generalization across shapes that are not explicitly
included in the training set. We also note that these results should be interpreted as generalization within a fixed
topological class. The present coordinate transformation assumes that the
physical domains can be mapped homeomorphically to the reference disk.
Domains with holes, disconnected components, or severe geometric
distortions would require additional treatment.

\begin{table}[t]
    \centering
    \caption{
    Mean relative $L_2$ errors of the zero-shot LTO predictions on unseen
    geometries.
    }
    \label{tab:unseen_geometry_errors}
    \begin{tabular}{lccc}
        \toprule
        &
        Circle
        &
        Quadrilateral
        &
        Pentagon
        \\
        \midrule
        $E$
        &
        $2.13\%$
        &
        $2.35\%$
        &
        $2.02\%$
        \\
        \bottomrule
    \end{tabular}
\end{table}

\subsubsection{Comparative study on the cost between numerical solver and PI-sampler}
To compare the computational cost and predictive performance of targeted
numerical data with those of the inexpensive physics-consistent data
generated by the PI-sampler for this problem, we conduct an additional
controlled experiment. Two LTOs are trained using data generated by a
finite-element solver and the PI-sampler, respectively. For the
finite-element baseline, we assume that the target forcing distributions
are known: the training forcings are sampled from the same distribution
families as the test forcings and boundary conditions reported in
Tables~\ref{tab:poisson_geometry_errors} and
\ref{tab:unseen_geometry_errors}, and the corresponding solutions are
computed using the finite-element-based PDE Toolbox in MATLAB. Further, we use the aforementioned BNN for generating the geometries in the training of these two cases. Thus, the
solver baseline has direct knowledge of the target forcing distributions.
In contrast, the PI-sampler does not use these distributions. Its training
pairs are obtained by sampling from the aforementioned collection of BNN
solution priors and applying the Poisson operator to obtain the corresponding
forcings. The two cases are evaluated using the same test data as those used in Tables~\ref{tab:poisson_geometry_errors} and
\ref{tab:unseen_geometry_errors}. 

The function encoder is pretrained exclusively using data generated by the PI-sampler and is subsequently kept fixed. When evaluated on the forcing and solution fields in both the solver-generated training dataset and the common test dataset, it achieves mean relative reconstruction errors of approximately \(1\%\). This result indicates that the pretrained encoder represents the fields from both data-generation procedures with comparable accuracy. We therefore use the same pretrained function encoder in both cases, together with identical Transformer architectures and optimization settings. Consequently, the comparison changes only the source of the operator-training pairs while holding the functional representation and downstream learning procedure fixed. For the computational comparison, the MATLAB finite-element solves are parallelized over 64 CPU cores on a system equipped with two Intel Xeon Platinum 8336C processors. The PI-sampler is evaluated both on the same CPU system and on an NVIDIA GeForce RTX 3090 GPU. The CPU-to-CPU results provide a hardware-matched comparison, while the GPU results illustrate the additional acceleration available through batched parallel generation.

\begin{figure}[h]
    \centering
    \includegraphics[width=\textwidth]{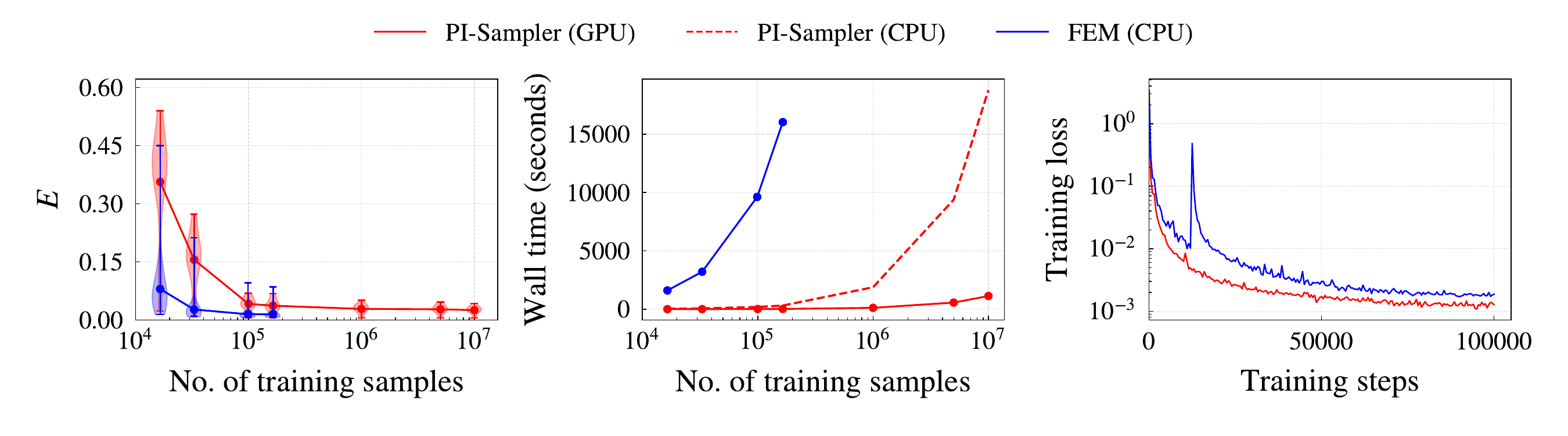}
    \caption{{Comparison of target-aware finite-element training
    data and target-agnostic PI-sampler data for the Poisson problem.
    Left: mean relative $L_2$ test error $E$ as a function of the number of
    training samples. Each marker is the mean over the test cases reported
    in Tables~\ref{tab:poisson_geometry_errors} and
    \ref{tab:unseen_geometry_errors}, and the error bars are the standard
    deviations across these test cases rather than uncertainty estimates
    from repeated training runs. Middle: wall-clock time required to
    generate the corresponding training datasets. Right: training-loss
    histories for Transformers trained with $10^5$ samples from the two
    data-generation procedures; the model architecture and training
    configuration are identical.}}
    \label{fig:poisson_data_comparison}
\end{figure}

The left panel of Fig. 5 reports the mean and standard deviation of the test errors for different training-set sizes. Three observations can be made. First, the mean errors obtained with both data-generation procedures generally decrease as the number of training samples increases. Second, the FEM-generated data attain lower mean errors with fewer samples, as expected, because their forcing-boundary distributions are explicitly matched to the target distributions. However, their errors exhibit larger standard deviations across the test cases. Third, the PI-sampler produces slightly higher mean errors but smaller cross-case dispersion, indicating more uniform, although less target-focused, performance over the tested forcing distributions.

A possible explanation for the difference lies in the sampling strategies. Specifically, the FEM training data are drawn from 33 prescribed forcing–boundary-condition configurations, whereas the PI-sampler generates training pairs using three different priors on the solution functions. For a fixed total number of training samples, the FEM approach allocates relatively fewer samples to each forcing–boundary-condition configuration, potentially leading to uneven approximation accuracy across the test cases. In particular, the largest errors obtained with the FEM-generated data occur primarily for test cases involving more rapidly fluctuating forcing functions, which are generally more challenging to approximate and may require more training samples to achieve comparable accuracy. In contrast, the PI-sampler generates samples from three implicitly induced forcing distributions, which may provide more balanced coverage of the tested cases and consequently reduce the variation in prediction errors. Nevertheless, this explanation remains qualitative, as the coverage of the induced forcing distributions is not explicitly controlled. Finally, because these standard deviations are computed across heterogeneous test cases, they quantify cross-case variability and should not be interpreted as uncertainty arising from random initialization or stochastic training.


Relevant remarks for the specific test are as follows:
\begin{itemize}
 \item In all other numerical experiments, the PI-sampler generates
    training pairs on the fly, which avoids storing a complete dataset and
    simplifies the training pipeline. For the present controlled comparison,
    however, the PI-sampler dataset is generated in advance and saved to
    disk so that its data-generation time can be compared directly with that of the finite-element solver under the same offline-dataset setting.
    
    \item The mean test errors with approximately $10^5$ training samples in these two cases are comparable as illustrated in the left panel of Fig. \ref{fig:poisson_data_comparison}.  For the computational cost comparison, generating approximately \(10^5\) training samples using MATLAB on the CPU requires \(9622.05\) seconds using the finite-element solver, compared with approximately \(187.78\) seconds for the PI-sampler on the same CPU system. The resulting hardware-matched speedup is approximately \(51.2\times\). When the PI-sampler is instead executed on an NVIDIA GeForce RTX 3090 GPU using batched generation, the time is reduced to \(11.29\) seconds, corresponding to an approximately \(852\times\) speedup relative to the CPU-based finite-element solver. We regard the CPU-to-CPU result as the primary hardware-matched comparison, while the GPU result illustrates the additional acceleration enabled by batched parallel generation.

    \item The PI-sampler generation is performed in batches on the CPU and GPU with 96 and 400
    samples generated per batch due to the available cores for CPU and GPU memory, respectively. A GPU
    with greater memory or throughput could permit larger batches and further
    reduce the wall-clock time.
    \item The right panel of Fig. ~\ref{fig:poisson_data_comparison} shows
    representative loss histories for the two Transformer-based operator models trained with $10^5$
    samples. In both cases, the training loss largely plateaus after
    approximately $10^5$ optimization steps. Because the Transformer
    architecture, batch size, and other training settings are identical,
    the training times are also nearly identical, at approximately three
    hours in the present experiments.
\end{itemize}

Overall, the solver-based training data are more
sample-efficient when the target distribution is known, whereas the main
advantage of the PI-sampler is the substantially lower cost of generating
large datasets. Solver-generated data may therefore be preferable when a
narrow target distribution is known and maximum accuracy per sample is
required, while the PI-sampler is attractive for broad data augmentation or general-purpose pretraining when downstream forcing distributions are not known in advance. The reported generation times exclude the subsequent training of the LTO.

\subsection{Time-dependent Sine--Gordon equation}
\label{sec:sine_gordon}

{The preceding examples are steady; we now test the framework on the two-dimensional time-dependent Sine--Gordon equation:}
\begin{equation}
\frac{\partial^2 u}{\partial t^2}
-
\Delta u
+
\sin(u)
=
f,
\qquad
\boldsymbol{x}
\in
[-1,1]^2,
\quad
t\in[0,1].
\label{eq:sine_gordon}
\end{equation}
Dirichlet boundary conditions are imposed on the spatial boundary:
\begin{equation}
u(t,\boldsymbol{x}_{\mathrm{bc}})
=
b(t,\boldsymbol{x}_{\mathrm{bc}}),
\qquad
\boldsymbol{x}_{\mathrm{bc}}
\in
\partial[-1,1]^2,
\label{eq:sine_gordon_bc}
\end{equation}
and the initial condition is
\begin{equation}
u(0,\boldsymbol{x})
=
u_0(\boldsymbol{x}),
\qquad
u_t(0, \boldsymbol{x})
= 
v_0(\boldsymbol{x}).
\label{eq:sine_gordon_ic}
\end{equation}
The LTO learns the mapping
\[
\mathcal{G}_{\mathrm{SG}}:
(f,b,u_0, v_0)
\longmapsto
u(t, \bm{x}),
\qquad
\boldsymbol{x}
\in
[-1,1]^2,
\quad
t\in[0,1].
\label{eq:sine_gordon_operator}
\]

To generate training data with different temporal and spatial
characteristics, nine anisotropic BNN prior configurations are employed.
The temporal and spatial first-layer weights use different prior scales, as
described in {\ref{app:data_generation}}.

For the in-distribution evaluation, 100 test samples are drawn from each
of the nine prior configurations, giving a total of 900 samples. Each
solution is evaluated on a uniform grid containing
$ x \times y \times t = 
40
\times
40
\times
40
=
64{,}000
$
space-time points. The mean relative $L_2$ error is
$
E_{\mathrm{ID}}
=
1.48\%.
\label{eq:sine_gordon_id_error}
$

{To move beyond the in-distribution setting, we also evaluate the LTO using forcing functions drawn independently}
from Gaussian processes with correlation lengths
\[
l_t
\in
\left\{
0.25,
0.3,
0.5
\right\},
\qquad
l_x
=
l_y
\in
\left\{
0.25,
0.3,
0.5
\right\}.
\label{eq:sine_gordon_test_lengths}
\]
A total of 100 forcing realizations are generated, and the corresponding
initial and boundary conditions are evaluated based on samples from the BNN priors used for generating the training data. We
also consider 100 cases with homogeneous forcing,
$
f
=
0,
$
paired with independently generated initial and boundary conditions from the BNN priors in the above test cases.
Reference solutions are obtained using a high-resolution finite-difference
method.
The mean relative errors are 
$
5.16\%,
$ and
$
5.35\%
$, respectively. 
The larger errors relative to the one-dimensional reaction--diffusion case
are expected because the Sine-Gordon equation involves nonlinear, 
time-dependent dynamics in a three-dimensional space-time domain. In
addition, the independent GP test distributions introduce combinations of
temporal and spatial length scales that differ from those induced by the
training priors.

{
Despite this distribution shift, the LTO maintains errors near $5\%$ on the
two prescribed test families. These results provide empirical evidence that
the sampled training data contain enough variability for these particular
tests beyond the exact BNN prior configurations. They also show that the same
sampling construction can be applied when the forcing is obtained through a
nonlinear operator containing both derivatives and $\sin(u)$.
}

\begin{table}[t]
    \centering
    \caption{
    Mean relative $L_2$ errors of the zero-shot LTO predictions for the
    time-dependent Sine--Gordon equation.
    }
    \label{tab:sine_gordon_errors}
    \begin{tabular}{lccc}
        \hline
        Test set
        &
        In-distribution
        &
        $f \sim \mathcal{GP}$
        &
        $f=0$
        \\
        \hline
        $E$ 
        &
        $1.48\%$
        &
        $5.16\%$
        &
        $5.35\%$
        \\
        \hline
    \end{tabular}
\end{table}

\subsection{High-dimensional nonstationary PDE}
\label{sec:high_dimensional_pde}

{We next examine whether the same pipeline remains computationally viable as the input dimension increases. For this purpose, we consider a five-dimensional nonstationary PDE adapted from}
Ref.~\cite{han2018solving}. The governing equation is

\begin{equation}
\frac{\partial u}{\partial t}
(t,\boldsymbol{x})
-
0.02
\sum_{i=1}^{d}
x_i^2
\frac{\partial^2 u}{\partial x_i^2}
(t,\boldsymbol{x})
+
0.02
\sum_{i=1}^{d}
x_i
\frac{\partial u}{\partial x_i}
(t,\boldsymbol{x})
=
f(t,\boldsymbol{x}),
\label{eq:high_dimensional_pde}
\end{equation}
where
\begin{equation}
d
=
5,
\qquad
\boldsymbol{x}
\in
B^5,
\qquad
t
\in
[0,1],
\label{eq:high_dimensional_domain}
\end{equation}
and $B^5$ denotes the five-dimensional unit ball. Homogeneous initial and
boundary conditions are imposed:
\[
u(0,\boldsymbol{x})
=
0,
\qquad
\boldsymbol{x}
\in
B^5,
\label{eq:high_dimensional_ic}
\]
and
\begin{equation}
u(t,\boldsymbol{x})
=
0,
\qquad
(t,\boldsymbol{x})
\in
[0,1]
\times
\partial B^5.
\label{eq:high_dimensional_bc}
\end{equation}
The homogeneous initial and boundary conditions are enforced in both the PI-sampler and the function encoder’s solution decoder by multiplying their respective unconstrained outputs by \(t(1-\|\bm{x}\|^2)\). The forcing is computed by applying the governing differential operator to the complete constrained solution field. The LTO learns the mapping
\[
\mathcal{G}_{5\mathrm{D}}:
f
\longmapsto
u.
\label{eq:high_dimensional_operator}
\]

Obtaining large numbers of independent reference solutions using
conventional grid-based numerical solvers is computationally prohibitive in
the six-dimensional space--time domain. We therefore evaluate the model
using 1000 in-distribution samples generated from the same BNN prior family
used for training. For each sample, the prediction error is computed on
10,000 randomly selected space--time points. The resulting mean relative error is
$
E_{5\mathrm{D}}
=
5.18\%.
$
{This result shows that the PI-sampler, function encoder, and LTO can be trained and evaluated for this five-dimensional example.}

{The evidential scope of this result is limited, however, because the experiment is an in-distribution representation test:}
both the training and testing functions are generated from the same BNN
prior family. It therefore demonstrates the ability of the LTO to learn the
operator over the sampled function class, but it does not by itself
establish generalization to arbitrary independently prescribed forcing
distributions. Independent manufactured-solution families or alternative
high-dimensional reference methods would provide a stronger
out-of-distribution assessment and are left for future study.

\subsection{{Vortex-induced vibration of a flexible cylinder}}
\label{sec:viv}

{
{
Vortex-induced vibration (VIV) of a flexible cylinder provides an
engineering test involving fluid--structure interaction
\cite{fan2019mapping,kharazmi2021inferring}. We focus on the structural response
in the cross-flow direction, illustrated in
Fig.~\ref{fig:viv_results}(a). Because the displacement equations have the
same form in the two transverse directions, we use the cross-flow displacement
$u(t,z)$ in the $y$-direction as the representative case:}
\begin{equation}
\frac{\partial^2 u}{\partial t^2}
+2\zeta\omega_n\frac{\partial u}{\partial t}
-\frac{T}{\mu}\frac{\partial^2 u}{\partial z^2}
=\frac{f_l}{\mu},
\qquad z\in[0,L], ~ t \in [0, 1],
\label{eq:viv_governing_equation}
\end{equation}
where $f_l(t,z)$ is the cross-flow hydrodynamic forcing in the {$y$-direction}, $\zeta$ is the
damping ratio, $\omega_n$ is the natural angular frequency, $T$ is the
structural tension, and $\mu$ is the cylinder mass per unit length. The
cylinder is pinned at both ends, so that $u(t,0)=u(t,L)=0$, which are hard encoded in the model similar as treatment of the boundary conditions in Sec. \ref{sec:high_dimensional_pde}. 
{The initial displacement and velocity are prescribed as $u(0,z)=u_0(z)$ and $u_t(0,z)=v_0(z)$ for $z\in[0,L]$. Accordingly, the operator-learning task is $(f_l,u_0,v_0)\mapsto u$ over a unit time window.}

The physical parameters follow the experimental configuration of
Fan et al.~\cite{fan2019mapping} and are listed in
Table~\ref{tab:viv_parameters}. Here $f_n=U/(U_r d)$,
$\omega_n=2\pi f_n$, and
$T=(f_nL)^2(4\mu+C_m\rho\pi d^2)$. The cylinder has aspect ratio
$L/d=240$, and $C_m=1$ is the uniform added-mass coefficient used to define
the first-mode natural frequency.

\begin{table}[H]
    \centering
    \caption{{
    Physical parameters used in the vortex-induced-vibration experiment.}}
    \label{tab:viv_parameters}
    \begin{tabular}{cccccccc}
        \hline
        $U$ & $\rho$ & $d$ & $U_r$ & $\zeta$ & $\mu$ & $L$ & $C_m$ \\
        \hline
        $1.0$ & $1.0$ & $1.0$ & $9.787$ & $0.087$ & $4.0$ & $240.0$ & $1.0$ \\
        \hline
    \end{tabular}
\end{table}


{With the physical model specified, we first evaluate the learned local-time operator.} The PI-sampler generates training pairs over the interval
$t\in[0,1]$ using the anisotropic BNN priors specified in {\ref{sec:data_sampler}. We}
first evaluate the trained LTO on 1280 new samples drawn from the same prior.
The mean relative $L_2$ error is $0.44\%$, showing that the latent
representation resolves the sampled displacement and forcing fields
accurately within the training regime.


{We then move from in-distribution unit-window tests to long-horizon reconstruction using the experimental cross-flow displacement measurements of
Fan et al.~\cite{fan2019mapping}, which span $t\in[0,40]$.} Because the LTO
is trained on intervals of length one, the full interval is partitioned into
40 windows $I_k=[k-1,k]$, $k=1,\ldots,40$, and the model is applied
recurrently. The terminal predicted state in each window supplies the
initial state for the next window.

For this evaluation, the forcing history is taken from  Fan et al. ~\cite{fan2019mapping}, and it is reported to be able to generate the displacement that agrees well with the experiment in \cite{fan2019mapping}. The pretrained LTO is then given this forcing and the recurrently
propagated initial states to reconstruct $u$. This choice retains forcing
patterns associated with the measured fluid-structure interaction, rather
than replacing them by an independently selected GP test distribution. The zero-shot recurrent prediction has a mean relative $L_2$ error of
$4.54\%$ over all 40 windows. Figure~\ref{fig:viv_results}(b) shows that the
predicted displacement remains close to the experimental reference despite
the repeated propagation of the predicted state.

{
We use the FSI-simulation forcing from \cite{fan2019mapping} rather than independently drawn GP samples because it preserves forcing patterns characteristic of the physical problem. The present experiment evaluates the structural response to a prescribed forcing history, without feedback from the predicted motion to the fluid. A stronger future test would integrate the LTO as a structural surrogate within a two-way coupled fluid–structure simulation.
}

\begin{figure}[H]
    \centering
    \begin{minipage}[b]{0.40\textwidth}
        \centering
        \begin{overpic}[width=\linewidth,height=4.2cm]{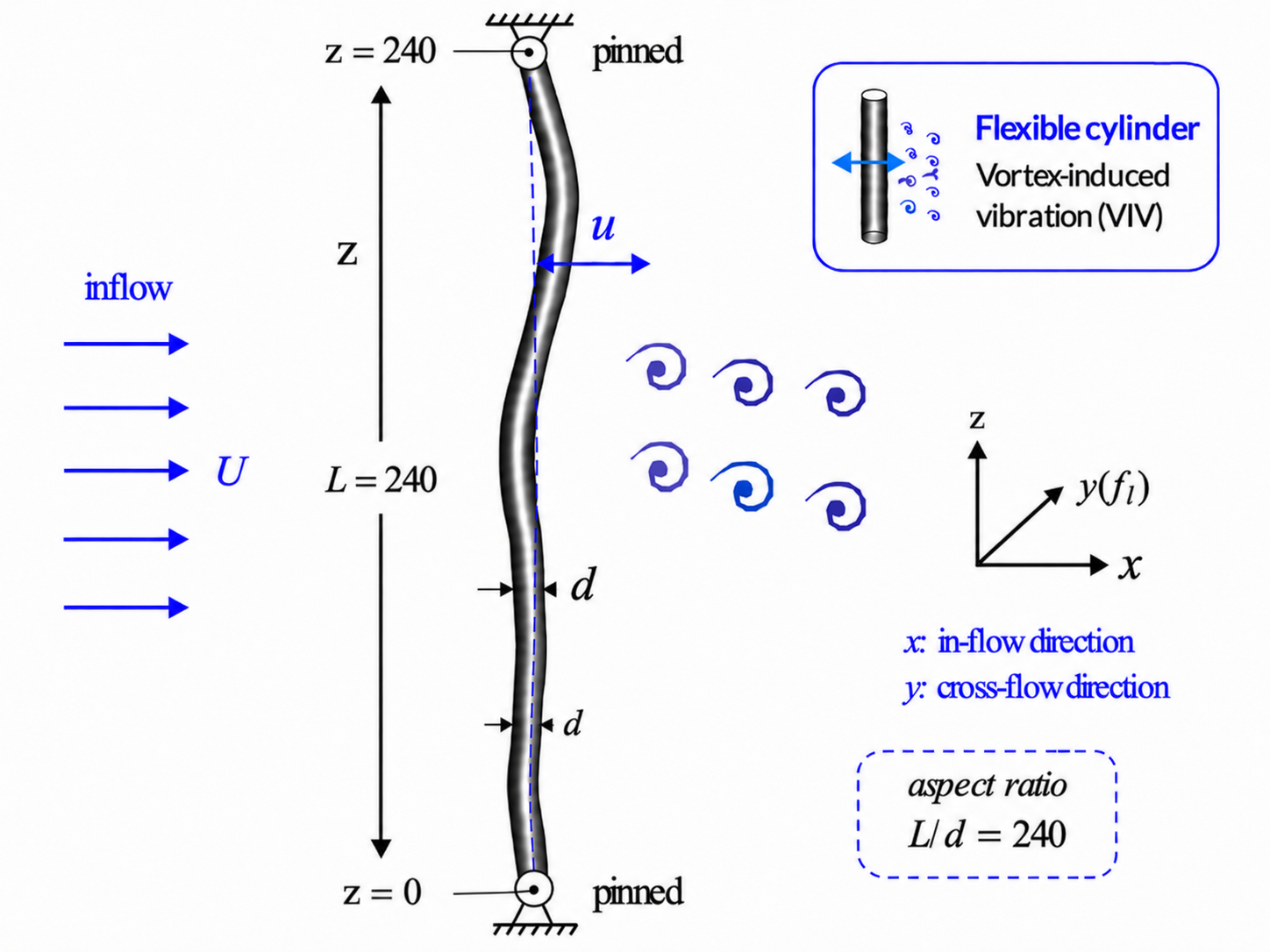}
            \put(5,75){%
                {\setlength{\fboxsep}{0.6pt}\colorbox{white}{\color{black}\bfseries a}}%
            }
        \end{overpic}
    \end{minipage}
    \hfill
    \begin{minipage}[b]{0.58\textwidth}
        \centering
        \begin{overpic}[width=\linewidth,height=4.2cm]{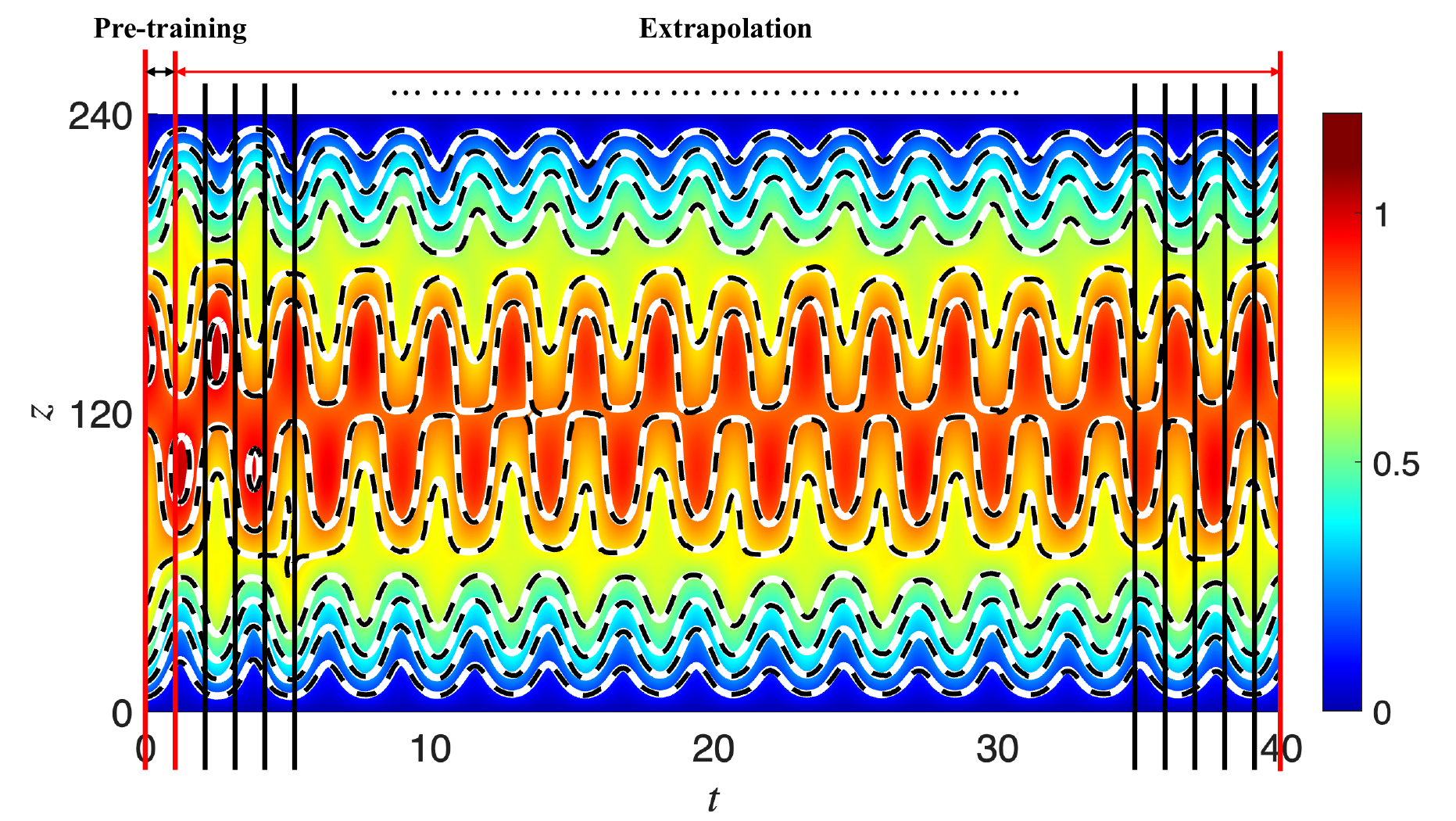}
            \put(1,50){%
                {\setlength{\fboxsep}{0.6pt}\colorbox{white}{\color{black}\bfseries b}}%
            }
        \end{overpic}
    \end{minipage}
    \caption{{
    Vortex-induced vibration of a flexible cylinder. (a) Schematic of the
    cylinder and the cross-flow displacement $u$ (in the $y-$direction); $\xi$ denotes the in-flow
    displacement. (b) Recurrent reconstruction over
    $t\in[0,40]$ using an LTO trained on unit-length time windows. The colored
    field and white curves show the experimental displacement, while the
    black dashed curves show the LTO prediction.}}
    \label{fig:viv_results}
\end{figure}



{Because recurrent application may accumulate error across windows, we next refine the LTO-predicted solution
coefficients independently in each window by minimizing}
the physics-informed loss in Sec. \ref{sec:lto}} for {100 L-BFGS iterations}. Only the latent
solution coefficients are optimized; the learned basis and Transformer
parameters remain fixed. This lightweight refinement reduces the mean
relative $L_2$ error from $4.54\%$ to $0.74\%$.

To test the effectiveness of the pretrained model for fine-tuning, we perform the following comparative study: (1) we keep the learned basis from the FE, and repeat the same coefficient-space optimization with all coefficients initialized to zero, and (2) we employ a vanilla PINN to solve the governing equation given the forcing, boundary, and initial conditions.  The formulation of loss, optimizer, training steps,
and recurrent procedure are kept the same as in the previous fine-tuning case. As summarized in
Table~\ref{tab:viv_errors}, the mean relative test errors for these two cases are $77.52\%$ and $97.88\%$, respectively. Details on the vanilla PINNs, e.g., architecture, activation, etc., are present in \ref{sec:computations}.

\begin{table}[H]
    \centering
    \caption{{
    Mean relative $L_2$ errors for long-horizon recurrent reconstruction of
    the vortex-induced-vibration displacement.}}
    \label{tab:viv_errors}
    \begin{tabular}{lc}
        \hline
        Method &$E$ \\
        \hline
        Zero-shot LTO & $4.54\%$ \\
        LTO initialization $+$ physics-informed refinement & $0.74\%$ \\
        Zero initialization $+$ physics-informed refinement & $77.52\%$ \\
        {Vanilla PINNs}      & $97.88\%$\\
        \hline
    \end{tabular}
\end{table}

For this VIV example, the governing equation and boundary and initial conditions are linear in the solution. With the basis and collocation points fixed, coefficient refinement is therefore a linear least-squares problem that could alternatively be solved directly. The FE training objective does not explicitly enforce orthogonality or linear independence of the learned basis, so the conditioning of the resulting system merits investigation. We use L-BFGS refinement initialized by the LTO because this formulation also accommodates nonlinear governing equations. The results demonstrate the benefit of LTO initialization within the prescribed iteration budget, but do not establish superiority over direct least-squares methods or fully converged coefficient optimization. A systematic comparison of solver choices and regularization is left for future work.

{

}

\clearpage
\section{Summary}
\label{sec:overall_discussion}
\label{sec:conclusion}

{
This work introduced a physics-informed sampler for generating
PDE-consistent training data for operator learning. Rather than first
prescribing source terms and conditions and then repeatedly solving the
PDE, the PI-sampler draws candidate solution fields from BNN priors and
applies the differential, boundary, and initial-condition operators through
automatic differentiation. The BNN prior thereby defines a probability
measure over candidate solutions, and the governing operators induce an
implicit pushforward distribution over the corresponding PDE inputs. An
explicit density is not required for sampling, even when a nonlinear
operator makes the induced distribution non-Gaussian and analytically
intractable.

The numerical experiments clarify both the usefulness and the scope of this
construction. Mixtures of BNN priors with different scales broaden the
 characteristic frequencies represented in the generated
solutions, while anisotropic mixtures accommodate different spatial and
temporal scales. The resulting data supported accurate operator predictions
for independently prescribed GP forcings, homogeneous forcing, mixed
boundary conditions, unseen geometries, time-dependent dynamics, and an
in-distribution five-dimensional problem. For the reaction-diffusion case, the SVD diagnostics show overlap in selected normalized projections and increasing nearest-sample shape distances as the testing GP correlation length decreases. These observations
establish empirical adequacy for the cases studied rather than theoretical
coverage of arbitrary forcing distributions.

The PI-sampler is independent of the downstream operator architecture. In
this study it was combined primarily with an LTO, whose function encoder
compresses continuous fields into learned basis coefficients and whose
Transformer maps heterogeneous PDE inputs to solution coefficients. This
latent formulation accommodates multiple functional inputs, varying
geometries, irregular observations, and high-dimensional space--time
domains. The reaction--diffusion experiment with DeepONet further indicates
that the generated data can also be used with another neural-operator
architecture. In the vortex-induced-vibration experiment, the LTO prediction
also provided an effective initialization for low-dimensional
physics-informed refinement. 

In the controlled Poisson equation test case, for the specific hardware, resolutions, and numerical solvers used here,
the PI-sampler generated paired data about 51 and 850 times faster than the
corresponding solver-based procedures when run on the CPU and GPU, respectively. These timings compare only the
data-generation stages and exclude downstream model training. Moreover,
when the desired forcing functions or their probability distribution are
known in advance, sampling them directly and solving the PDE numerically can
produce a more targeted training dataset. The reported costs should
therefore be interpreted as problem-specific reference values, rather than
as a like-for-like comparison of identical training distributions or a
claim that PI-sampling is always preferable.

Several limitations remain. The practical performance depends on the
probability mass assigned by the BNN prior to physically relevant solution
regimes; broad topological support alone does not ensure useful finite-sample
coverage. Further, only continuous, smooth solutions are considered in the present work. Problems involving shocks, discontinuities, interfaces, or localized multiscale structures may require different generators, activation
functions, domain decompositions, or piecewise representations. Nonlinear
operators can also distort the solution prior into a highly nonuniform
forcing distribution, for which a general concentration analysis is not
provided here. In addition, the geometry experiment is limited to domains
that can be mapped to a common reference domain, the five-dimensional case
is evaluated only in distribution, and coefficient recovery from irregular
observations introduces an optimization whose stability depends on the
number, locations, and noise level of the observations. Controlled comparisons of multiple neural operators trained on the same generated dataset would also help isolate the effect of the data-generation strategy from that of the operator architecture.

These limitations motivate a complementary rather than replacement view of
the PI-sampler. When a target input distribution is known and high-fidelity
accuracy in particular regimes is required, solver-generated data can
anchor the training set. Larger numbers of inexpensive PI-sampler
realizations can then augment those data and broaden the represented
solution behaviors. Future work should investigate such hybrid or
multifidelity datasets, adaptive selection of prior components in poorly
represented regions, and more expressive stochastic generators for
non-Gaussian or nonsmooth solution classes. Furthermore, as aforementioned, the PI-sampler can be viewed as an implicit generative model. Related
physics-informed generative frameworks~\cite{yang2020physics,zhou2026scalable}, have been developed for forward and inverse stochastic differential equations. Motivated by these
developments, another promising direction is to tune the BNN priors so that
the empirical distribution of the induced forcing
$f=\mathcal{N}_{\bm{\beta}}[u]$ approximately matches a specified target
distribution. Such distribution-guided prior calibration could make the
generated data more targeted while retaining the efficient sampling procedure
of the PI-sampler.

The potential value of PI-sampling extends beyond data generation for a single prescribed input distribution. Pretraining general-purpose PDE models may require datasets spanning multiple equations, physical parameters, geometries, and functional scales, for which exclusive reliance on repeated numerical solves can impose substantial computational costs. PI-sampling offers an inexpensive source of continuous, PDE-consistent training pairs that could complement solver-generated datasets in this setting. Its value for such pretraining will depend on whether the induced solution and input distributions capture behaviors useful for transfer to downstream problems, rather than on dataset size alone.

In summary, the PI-sampler efficiently generates diverse, PDE-consistent
training pairs. Its induced distribution must be assessed for each target
problem, and the method is best viewed as physics-informed augmentation of
targeted high-fidelity simulations.
}


\section*{Acknowledgements}
W. Y., S. Z., and X. M. acknowledge the support of the National Natural Science Foundation of China (No. 12201229) and the Interdisciplinary Research Program of HUST (No. 2024JCYJ003). X. M. also acknowledges the support of the Xiaomi Young Talents Program.  
\appendix

\section{Latent Transformer operator details}
\label{sec:lto_details}

\subsection{ Functional representations and coefficient recovery}

For a functional input or output \(\gamma\), the coefficient network receives the sampled representation
\[
U_\gamma
=
[\gamma(\bm{\xi}_1),\ldots,\gamma(\bm{\xi}_{N_\xi})]^T.
\]
Its coefficients and continuous reconstruction are
\[
\psi_\gamma
=
\mathrm{NN}_{C,\gamma}(U_\gamma;\bm{\theta}_{C,\gamma}),
\qquad
\gamma_{\mathrm{FE}}(\bm{\xi})
=
\psi_\gamma^T
\mathrm{NN}_{B}(F(\bm{\xi});\bm{\theta}_B).
\]
Separate coefficient networks are used for the source and solution fields. A common basis may be used when suitable; alternatively, distinct source and solution bases can accommodate differences in regularity or imposed conditions. The basis notation is suppressed in the main text when the represented field is clear.

This decomposition is closely related to the branch–trunk representation of DeepONet \cite{lu2021learning}. Its role here is to construct a representation of each field before learning the mapping between PDE inputs and solutions.

When observations are available on the training point set, the pretrained coefficient network is used directly. For a different observation set
\[
\mathcal S_{\mathrm{obs}}
=
\{(\widetilde{\bm{\xi}}_j,\gamma_{\mathrm{obs}}(\widetilde{\bm{\xi}}_j))\}_{j=1}^{N_{\mathrm{obs}}},
\]
the coefficients are recovered by
\begin{equation}
\widehat{\psi}_\gamma
=
\underset{\psi\in\mathbb R^p}{\arg\min}
\left\{
\frac1{N_{\mathrm{obs}}}
\sum_{j=1}^{N_{\mathrm{obs}}}
\left|
\gamma_{\mathrm{obs}}(\widetilde{\bm{\xi}}_j)
-
\psi^T\phi(\widetilde{\bm{\xi}}_j)
\right|^2
+
\lambda_\psi\|\psi\|_2^2
\right\}.
\end{equation}
In the present study, (1) \(\lambda_\psi=0\) unless otherwise stated, and (2) the coefficients are recovered via the gradient descent approach with a fixed basis.


A standard DeepONet branch network typically expects observations at prescribed sensors. The present recovery procedure instead fits coefficients in a continuous pretrained basis, allowing different observation sets without retraining that basis. This flexibility carries the additional cost of coefficient recovery and does not guarantee accurate reconstruction from arbitrarily sparse observations.

\subsection{Input tokens and Transformer mapping}

Let \(\Lambda\) collect the physical inputs relevant to a particular problem. Functional inputs are represented by FE coefficients or prescribed observations, while parameters and geometry use problem-specific representations. Each representation is linearly projected to the common embedding dimension \(d_{\mathrm{emb}}\).

A representative initial token sequence is
\[
Z^{(0)}
=
[z_f,z_b,z_{u_0},z_{v_0},z_\beta,z_\Omega,z_{\mathrm{out}}^{(0)}],
\]
where only the modalities required by the PDE are retained. The temporal derivative token can also be included for second-order time-dependent problems. The learnable token \(z_{\mathrm{out}}^{(0)}\) aggregates information for predicting the solution coefficients. Extra type embeddings are employed to identify different physical modalities.

When a coefficient vector is partitioned into multiple tokens, we write
\[
\psi_\gamma
=
\operatorname{concat}
\left(
\psi_\gamma^{(1)},\ldots,\psi_\gamma^{(N_\gamma)}
\right).
\]
Each part is projected separately, and positional embeddings may identify different parts.

The token sequence is processed by \(L_T\) Transformer blocks:
\[
\widetilde Z^{(\ell)}
=
Z^{(\ell-1)}
+
\operatorname{MSA}
\left(
\operatorname{LN}(Z^{(\ell-1)})
\right),
\]
\[
Z^{(\ell)}
=
\widetilde Z^{(\ell)}
+
\operatorname{MLP}
\left(
\operatorname{LN}(\widetilde Z^{(\ell)})
\right),
\qquad \ell=1,\ldots,L_T.
\]
Here, MSA denotes multi-head self-attention, LN denotes layer normalization, and MLP denotes the feed-forward subnetwork. The feed-forward sublayers use GELU activations.

The final, input-dependent representation of the output token is projected to the predicted solution coefficients:
\[
\widehat{\psi}_u
=
W_{\mathrm{out}}z_{\mathrm{out}}^{(L_T)}
+b_{\mathrm{out}}.
\]

The coefficient loss in Sec. \ref{sec:lto} may be supplemented by a field-space reconstruction loss,
\begin{equation}
\mathcal L_u
=
\frac1{B_{\mathrm{OL}}N_\xi}
\sum_{k=1}^{B_{\mathrm{OL}}}
\sum_{j=1}^{N_\xi}
\left|
\widehat u^{(k)}({\bm{\xi}}_j)-u^{(k)}({\bm{\xi}}_j)
\right|^2,
\end{equation}
giving
\[
\mathcal L_{\mathrm{LTO}}
=
\mathcal L_{\mathrm{OL}}+\lambda_u\mathcal L_u.
\]
The field-space term accounts directly for reconstruction error in a potentially nonorthogonal learned basis. In the present study, \(\lambda_u=0\) unless otherwise stated.

\subsection{Physics-informed coefficient refinement}

For a fixed test instance, the residual and boundary losses in Sec. \ref{sec:lto} are
\[
\mathcal L_{\mathrm{eq}}
=
\frac1{N_r}
\sum_{j=1}^{N_r}
\left|
\mathcal N_\beta[u_\psi]({\bm{\xi}}_j^r)
-f({\bm{\xi}}_j^r)
\right|^2,
\]
\[
\mathcal L_{\mathrm{bc}}
=
\frac1{N_{\mathrm{bc}}}
\sum_{j=1}^{N_{\mathrm{bc}}}
\left|
\mathcal B_\beta[u_\psi]({\bm{\xi}}_j^{\mathrm{bc}})
-b({\bm{\xi}}_j^{\mathrm{bc}})
\right|^2.
\]

For initial displacement data, define
\[
\mathcal L_{u_0}
=
\frac1{N_{\mathrm{ic}}}
\sum_{j=1}^{N_{\mathrm{ic}}}
\left|
u_\psi(0,\bm{x}_j^{\mathrm{ic}})
-u_0(\bm{x}_j^{\mathrm{ic}})
\right|^2.
\]
For second-order time-dependent problems, also define
\[
\mathcal L_{v_0}
=
\frac1{N_{\mathrm{ic}}}
\sum_{j=1}^{N_{\mathrm{ic}}}
\left|
\partial_tu_\psi(0,{\bm{x}_j}^{\mathrm{ic}})
-v_0({\bm{x}_j}^{\mathrm{ic}})
\right|^2.
\]
The initial-data loss is
\[
\mathcal L_{\mathrm{ic}}
=
\begin{cases}
\mathcal L_{u_0},
& \text{for first-order time-dependent problems},\\[2mm]
\mathcal L_{u_0}+\eta_v\mathcal L_{v_0},
& \text{for second-order time-dependent problems},
\end{cases}
\]
where \(\eta_v\) controls the relative weighting of temporal derivative data. The initial-data term is omitted for steady problems. Any condition enforced exactly by the representation can likewise be omitted from the penalty.

Only the solution coefficients are optimized, starting from the LTO prediction. For linear governing and constraint operators with fixed collocation points, the resulting objective is a linear least-squares objective. For nonlinear governing operators, it generally becomes a nonlinear least-squares objective. The VIV experiment evaluates the benefit of LTO initialization under the specified iterative optimization budget; it does not compare against a direct least-squares solution.

\section{PI-sampler configurations}
\label{app:data_generation}
\label{sec:data_sampler}

{
This appendix reports the BNN priors used by the PI-sampler in the five
numerical examples. In each case, the solution generator is a one-hidden-layer
BNN with Fourier features. If the hidden-layer width is $H$, the output weights
have standard deviation $1/\sqrt{H}$. The first-layer standard deviations
control the characteristic variation of the sampled solution fields and are
listed in Table~\ref{tab:pi_sampler_priors}. For time-dependent problems,
separate temporal and spatial values are used when the two directions have
different characteristic scales.  For each training sample, one of the \(M\) prior configurations is selected uniformly at random, after which the network parameters are sampled from the selected configuration. The mixture weights are therefore \(\pi_m=1/M\), \(m=1,\ldots,M\).

\begin{table}[H]
\caption{{BNN-prior configurations used by the PI-sampler. The
listed values are standard deviations of the first-layer weights.}}
\label{tab:pi_sampler_priors}
\centering
\scriptsize
\renewcommand{\arraystretch}{1.18}
\setlength{\tabcolsep}{3.5pt}
\begin{tabularx}{\textwidth}{l c >{\raggedright\arraybackslash}X}
\toprule
Case & $H$ & First-layer prior configuration \\
\midrule
Reaction--diffusion (Sec.~\ref{sec:reaction_diffusion})
& 100 & Isotropic standard deviation in
$\{8,10,12,14,18,20\}$; the additional values $\{2,4,6\}$ are used for
the homogeneous and low-frequency cases. \\
\makecell[l]{Poisson equation  on \\parameterized geometries} (Sec.~\ref{sec:poisson_geometry})
& 256 & Isotropic standard deviation in $\{3,5,7\}$. The auxiliary BNN used
to generate the geometry has first-layer standard deviation $3$. \\
Sine--Gordon equation (Sec.~\ref{sec:sine_gordon})
& 100 & The temporal and spatial standard deviations are selected independently
from $\{2,4,5\}$, giving nine anisotropic configurations. \\
High-dimensional PDE (Sec.~\ref{sec:high_dimensional_pde})
& 256 & Isotropic standard deviation $2$ in all temporal and spatial directions. \\
Vortex-induced vibration (Sec.~\ref{sec:viv})
& 100 & Anisotropic standard deviations
$\sigma_t=1.25/t_{\mathrm{len}}$ and $\sigma_z=6/z_{\mathrm{len}}$, where
$t_{\mathrm{len}}=1$ and $z_{\mathrm{len}}=240$. \\
\bottomrule
\end{tabularx}
\end{table}

These configurations were selected empirically to generate solution fields
with the spatial and temporal scales relevant to each example. They define the
distribution induced by the PI-sampler; they are not claimed to reproduce a
prescribed closed-form distribution of the forcing term. When such a target
distribution is known explicitly, drawing the forcing term from it and solving
the PDE can provide more targeted training data. The PI-sampler instead serves
as a solver-free option when broad, continuously generated training data are
desired and an exact forcing distribution is unavailable or unnecessarily
restrictive.
}

\section{Implementation details}
\label{app:implementation_details}
\label{sec:computations}

{
All trainable networks are optimized with Adam and a cosine-annealing learning-rate
schedule. The function encoder (FE) represents a field using coefficients
produced by $\mathcal{NN}_C$ and basis functions produced by
$\mathcal{NN}_B$. Its architectures and training settings are summarized in
Tables~\ref{tab:fe_architecture} and~\ref{tab:discretization}. In the latent
Transformer operator (LTO), the FE coefficients and the sampled boundary or
initial data are embedded as tokens, mapped to a common model dimension, and
concatenated before the self-attention blocks. Table~\ref{tab:lto_architecture}
reports the corresponding LTO settings.

\begin{table}[H]
\caption{{Function-encoder architectures and optimization settings.}}
\label{tab:fe_architecture}
\centering
\scriptsize
\renewcommand{\arraystretch}{1.18}
\setlength{\tabcolsep}{3.2pt}
\begin{tabular}{l c c c c r}
\toprule
& \multicolumn{2}{c}{$\mathcal{NN}_C$} &
\multicolumn{2}{c}{$\mathcal{NN}_B$} & \\
\cmidrule(lr){2-3}\cmidrule(lr){4-5}
Case & Width $\times$ depth & Activation & Width $\times$ depth & Activation & Steps \\
\midrule
Reaction--diffusion & $256\times5$ & ReLU & $256\times4$ & tanh & 300{,}000 \\
Poisson geometry & $512\times5$ & ReLU & $512\times5$ & tanh & 200{,}000 \\
Sine--Gordon & $2048\times8$ & ReLU & $2048\times8$ & tanh & 500{,}000 \\
High-dimensional PDE & $1024\times8$ & ReLU & $1024\times8$ & tanh & 300{,}000 \\
Vortex-induced vibration & CNN & ReLU & $512\times4$ & tanh & 300{,}000 \\
\bottomrule
\end{tabular}
\end{table}

\begin{table}[H]
\caption{{Batch sizes and spatial--temporal discretizations used by
the function encoder and for evaluation.}}
\label{tab:discretization}
\centering
\scriptsize
\renewcommand{\arraystretch}{1.18}
\setlength{\tabcolsep}{3pt}
\begin{tabularx}{\textwidth}{l c >{\raggedright\arraybackslash}X >{\raggedright\arraybackslash}X}
\toprule
Case & Batch & FE training points per field & Evaluation points per field \\
\midrule
Reaction--diffusion & 600
& 256 equidistant points in $[-1,1]$
& 256 equidistant points in $[-1,1]$ \\
Poisson geometry & 768
& 1{,}000 points uniformly sampled in the unit disk
& Nodes of the MATLAB PDE Toolbox mesh on each geometry
(approximately 24{,}000--66{,}000) \\
Sine--Gordon & 144
& $30\times30\times30$ uniform tensor grid on
$[0,1]\times[-1,1]^2$
& $40\times40\times40$ uniform tensor grid on
$[0,1]\times[-1,1]^2$ \\
High-dimensional PDE & 64
& 5{,}000 points uniformly sampled in the unit ball
& 10{,}000 points uniformly sampled in the unit ball \\
Vortex-induced vibration & 256
& $21\times256$ uniform grid on $[0,1]\times[0,240]$
& $801\times256$ uniform grid on $[0,40]\times[0,240]$ \\
\bottomrule
\end{tabularx}
\end{table}

\begin{table}[H]
\caption{{Latent Transformer operator architectures and training settings.}}
\label{tab:lto_architecture}
\centering
\scriptsize
\renewcommand{\arraystretch}{1.18}
\setlength{\tabcolsep}{4.1pt}
\begin{tabular}{l c c c c c r}
\toprule
Case & Model dim. & Blocks & Heads & FFN dim. & Batch & Steps \\
\midrule
Reaction--diffusion & 512 & 6 & 8 & 2048 & 600 & 200{,}000 \\
Poisson geometry & 256 & 4 & 8 & 1024 & 1{,}200 & 200{,}000 \\
Sine--Gordon & 1024 & 6 & 8 & 4096 & 144 & 300{,}000 \\
High-dimensional PDE & 1024 & 12 & 8 & 4096 & 128 & 300{,}000 \\
Vortex-induced vibration & 512 & 6 & 8 & 2048 & 256 & 1{,}000{,}000 \\
\bottomrule
\end{tabular}
\end{table}

The feed-forward sublayers of all Transformers use GELU activations. For the
reference solutions in the Sine--Gordon example, the finite-difference solver
uses a five-point stencil in space and a central-difference scheme in time on a
$400\times200\times200$ uniform grid in $(t,x,y)$.
}

{In the VIV test case, the employed PINN has a fully connected neural network to approximate $u_{\theta}(t,z)$, which consists of four hidden layers with 128 neurons per layer and uses the hyperbolic tangent activation function. At each optimization step, 4096 residual points, 512 boundary points, and 512 initial-condition points are sampled. The network is trained using the  L-BFGS for 100 iterations in each time window.}

In Sec. \ref{sec:poisson_geometry}, we employ the PDE Toolbox in MATLAB to solve the Poisson equation with diffusion coefficient $k=0.01$ and Dirichlet boundary conditions. The domains are discretized using unstructured triangular meshes with the maximum mesh-size parameter set to $0.01$. For the randomly generated domains in the validation set, the meshes contain 10,220--33,889 triangular elements. The circular, square, and pentagonal domains contain 27,978, 15,872, and 19,063 elements, respectively. The numerical solutions are then interpolated to the sampling points used by the operator network.

\bibliographystyle{elsarticle-num}
\bibliography{refs}
\end{document}